\documentclass[preprint,review,10pt,3p]{elsarticle}

\usepackage{amssymb}

\usepackage{lineno}

\usepackage{lipsum}
\usepackage{amsmath} 
\usepackage{relsize} 
\usepackage{commath}
\usepackage{hyperref}
\usepackage{tabularx}
\usepackage{booktabs}
\usepackage{multicol}
\usepackage{makecell}
\usepackage{subfig}
\usepackage{cleveref}
\usepackage{xcolor}
\newcommand{\hlcolor}{black}

\newcolumntype{Y}{>{\centering\arraybackslash}X}

\journal{Energy and Buildings}

\begin{document}

\begin{frontmatter}



\title{Impact of data for forecasting on performance of model predictive control in buildings with smart energy storage}

\author[EECi]{Max Langtry\corref{cor1}} \ead{mal84@cam.ac.uk}
\author[EECi]{Vijja Wichitwechkarn}
\author[EECi,ATI]{Rebecca Ward}
\author[EECi,ATI]{Chaoqun Zhuang}
\author[EECi,Surrey]{Monika J. Kreitmair}
\author[EECi,Surrey]{Nikolas Makasis}
\author[EECi,ATI]{Zack Xuereb Conti}
\author[EECi,ATI]{Ruchi Choudhary}

\cortext[cor1]{Corresponding author}

\affiliation[EECi]{organization={Energy Efficient Cities Initiative, Department of Engineering, University of Cambridge},
            addressline={Trumpington Street},
            city={Cambridge},
            postcode={CB2 1PZ},
            country={UK}}

\affiliation[ATI]{organization={Data-Centric Engineering, The Alan Turing Institute},
            addressline={British Library},
            city={London},
            postcode={NW1 2DB},
            country={UK}}

\affiliation[Surrey]{organization={School of Sustainability, Civil \& Environmental Engineering, University of Surrey},
            city={Guilford},
            postcode={GU2 7XH},
            country={UK}}

\begin{abstract}
Data is required to develop forecasting models for use in Model Predictive Control (MPC) schemes in building energy systems. However, data {\color{\hlcolor}is costly to both collect and exploit}. Determining cost optimal data usage {\color{\hlcolor}strategies} requires understanding of the forecast accuracy and resulting MPC operational performance it enables. This study investigates the performance of both simple and state-of-the-art machine learning prediction models for MPC in multi-building energy systems {\color{\hlcolor}using a simulated case study with} historic building energy data. The impact {\color{\hlcolor}on forecast accuracy of measures to improve model data efficiency are quantified, specifically for}: reuse of prediction models, reduction of training data {\color{\hlcolor}duration}, reduction of model data features, and online model training. A simple linear multi-layer perceptron model is shown to provide equivalent forecast accuracy to state-of-the-art models, with greater data efficiency and generalisability. The use of more than 2 years of training data for load prediction models provided no significant improvement in forecast accuracy. Forecast accuracy and data efficiency were improved simultaneously by using change-point analysis to screen training data. Reused models and those trained with 3 months of data had on average 10\% higher error than baseline, indicating that deploying MPC systems without prior data collection may be economic.
\end{abstract}


\begin{highlights}
\item Forecasting models for MPC tested using energy system simulation with historic data
\item Impact of data on forecast accuracy and MPC operational performance quantified
\item Simple linear MLP model provides equivalent accuracy to state-of-the-art models
\item More than 2 years of training data did not significantly improve forecast accuracy
\item Screening training data using change-points improves accuracy and data efficiency
\end{highlights}


\end{frontmatter}



\section{Introduction} \label{sec:intro}


\subsection{Background}

The operation of building systems accounts for 30\% of final energy consumption and 26\% of energy-related carbon emissions globally \cite{iea2023TrackingCleanEnergy}. Thus, decarbonizing building energy usage is necessary to achieve the targets of net-zero carbon emissions by 2050 \cite{committeeonclimatechange2020SixthCarbonBudget}. The use of distributed generation and storage technologies in building energy systems can enable substantial reductions in operational emissions \cite{aminitoosi2022BuildingDecarbonizationAssessing,oshaughnessey2021DemandSideOpportunityRoles,zhu2023ReviewDistributedEnergy}. Additionally, smart energy storage systems can reduce the impact of building energy usage on the electrical grid by allowing energy flexibility through demand-side management and demand response \cite{vazquez-canteli2020MARLISAMultiAgentReinforcement}, which is of particular importance in light of the expected electrification of heating demands \cite{leibowicz2018OptimalDecarbonizationPathways}. The effectiveness of these systems, quantified by operational performance metrics of the building energy system such as total electricity cost, incurred carbon emissions, and measures of grid impact, is determined by the ability of the storage control scheme to arbitrage energy and alter the timing of net energy usage to meet these operational goals. Hence, the development of performant storage control strategies has received substantial research attention in recent years \cite{drgona2020AllYouNeed,wang2020ReinforcementLearningBuilding,kathirgamanathan2021DatadrivenPredictiveControl}.

Model Predictive Control (MPC) is a prominent control methodology in the building energy systems literature. Its application to smart energy storage systems in buildings has been widely studied \cite{zhou2021IncorporatingDeepLearning,deng2023EvaluationDeployingDatadriven,souto2018ScenarioBasedDecentralizedMPC,touretzky2014IntegratingSchedulingControl,lee2020ModelPredictiveControl}, and has been found to provide substantial operational performance improvements compared to Rule-Based control (RBC) \cite{lee2020ModelPredictiveControl,erfani2021AnalysisImpactPredictive,oldewurtel2012UseModelPredictive}, the predominant technology used in existing real-world systems. MPC is shown to yield equivalent performance to Reinforcement Learning control (RLC) \cite{zhan2023ComparingModelPredictive}, the other major control methodology studied in current literature. Review papers \cite{wang2022ScienceMappingApproach,thieblemont2017PredictiveControlStrategies,farrokhifar2021ModelPredictiveControl} provide a thorough overview of the applications of MPC to building energy systems with distributed generation and storage.

MPC requires a model of the system dynamics, which estimates the next state of the system given; (a) the current system state, (b) current operational conditions, and (c) the applied control actions. Using this dynamics model and forecasts of the operational conditions over a given planning horizon, it predicts the operation of the system over that planning horizon. Optimization techniques are then used to determine the set of control actions that optimize a specified objective over the planning horizon. The performance of MPC therefore depends on: the accuracy of the system dynamics model, the accuracy of the operational condition forecasts, the ability of the optimization method to identify near-optimal control actions, and the match between the objective over the planning horizon and the global operational goals of energy management for the system. This work focuses on the accuracy of operational condition forecasts, and its resulting impact on the operational performance of MPC.

\subsection{Forecasting models for MPC}


A broad range of time series forecasting methodologies have been studied for the prediction of operational conditions for building energy management \cite{sun2020ReviewThestateoftheartDatadriven}, however recent literature has focused on the use of machine learning based methods \cite{zhang2021ReviewMachineLearning,chou2018ForecastingEnergyConsumption,dai2023ComparisonDifferentDeep,rahman2018PredictingElectricityConsumption,fan2019AssessmentDeepRecurrent} due to their promising performance. Successes in other fields have prompted the investigation of large-scale, high-complexity model architectures \cite{choi2023PerformanceEvaluationDeep,dai2023CityTFTTemporalFusion}. However, doubt remains as to whether such complex models are appropriate for use in the context of MPC for building energy systems \cite{zeng2023AreTransformersEffective,bunning2022PhysicsinformedLinearRegression,bunning2021ComparingMachineLearning}, due to computational and data availability limitations in practical systems. As such, this work considers both simple and high-complexity, state-of-the-art machine learning forecasting models.

\subsection{Study of impact of data on forecast accuracy and MPC performance}


As machine learning methods are purely data-driven, ``black box'' prediction models, achieving accurate forecasts requires the availability of training data that is representative of the building energy system for which the model will be used. However, acquiring representative training data is both challenging and costly. {\color{\hlcolor}The costs of acquiring data include not only the capital costs of installing monitoring systems, but also the costs of digital infrastructure, data processing, system maintenance, and quality assurance required to support data collection. For the case of developing forecasting models for MPC considered in this study, there are additional project costs associated with delaying the installation of the battery system whilst data is gathered. \cite{motegi2003CaseStudiesEnergy} estimates the capital cost of electricity and gas smart-metering for a university campus to be \$0.27/m$^2$, plus an additional \$0.11/m$^2$ for maintenance and supporting IT systems. However, this considers only the cost of collecting data, and neglects the significant costs of training and deploying machine learning models \cite{strubell2020EnergyPolicyConsiderations}, from both the computing and expertise required.}
Whilst the importance of data availability and the impact of data on forecast accuracy are widely acknowledged in the literature \cite{kathirgamanathan2021DatadrivenPredictiveControl,lee2020ModelPredictiveControl,wang2022ScienceMappingApproach,choi2023PerformanceEvaluationDeep,zhan2021DataRequirementsPerformance}, few works study the role of data in enabling good operational performance for MPC in building energy systems.

Determining cost optimal data collection strategies to support the development of forecasting models for MPC in buildings requires understanding of the trade-off between {\color{\hlcolor}the quantity of data \& data features used for model training}, and its associated costs, and the operational performance achieved by the controller. This impact of data for forecasting on MPC operational performance can be considered in two stages, by firstly studying the relationship between data and forecast accuracy of the resulting models, and then the sensitivity of the controller operational performance to the accuracy of forecasts.


Only a single previous study investigating the impact of data on the accuracy of forecasting models for building energy management could be identified. This work, \cite{choi2023PerformanceEvaluationDeep}, compares the prediction accuracy of deep learning architectures for forecasting thermal loads and building zone temperatures over varying training dataset sizes. It finds that increasing the training data {\color{\hlcolor}length} does not always improve prediction accuracy due to the strong seasonality of building energy behaviours, but that the addition of data with good similarity to the test dataset into the training dataset greatly improves forecast accuracy. A limitation of this study is that it analyses only a single year of thermal load and zone temperature data, synthesised from a building energy model and historical weather data, meaning that a limited range of training dataset sizes, from 3 to 9 months, is considered. As a result, training data preceding the test data by one year, which due to seasonality is likely to have the highest similarity and so greatest value, is unavailable, meaning the study of the benefit of additional training data is incomplete.


The impact of forecast accuracy on MPC operational performance has been studied to a limited extent. \cite{erfani2021AnalysisImpactPredictive} compares the use of various classical and machine learning forecasting models in a common MPC framework, quantifying both the forecast accuracy and resulting operational performance of MPC. However, the models studied all achieve comparable forecast accuracies and operational performances, meaning limited insight can be gained into the relationship between the two.
Further, as comparison is made between model types, these results cannot be used to assess the benefit of forecast improvement for any individual model. \cite{oldewurtel2012UseModelPredictive} shows that the use of more accurate, external weather forecasts improves MPC operational performance, but does not quantify the forecast accuracies for comparison. In \cite{bartolucci2019HybridRenewableEnergy}, the prediction accuracy and corresponding operational performance of MPC for two forecasts of electrical load with different levels of synthetic noise are quantified. Finally, \cite{enriquez2016SolarForecastingRequirements} provides the most complete study, investigating the variation of operational performance with the noise amplitude of synthetic forecasts of temperature and solar irradiance. However, as the forecast accuracy of the synthetic predictions is not quantified, these results cannot be compared to the performance of practical prediction models and used to assess the benefits of forecast accuracy improvements.


The direct impact of training data {\color{\hlcolor}length} on operational performance is studied in \cite{savadkoohi2023FacilitatingImplementationNeural}. It computes the operational performance achieved by a neural network based building thermal controller as the size of the training dataset varies, and finds that negligible performance improvements are obtained when using more than 8 months of training data. Whilst a comprehensive quantification of the relationship between training data volume and operational performance achieved by the specific control scheme studied is provided, the results are specific to the atypical controller architecture used, which does not include the explicit forecasting of operational variables used in typical MPC schemes. Additionally, other aspects of data, such as the inclusion of additional data variables, and the reuse of data from existing buildings, are not considered.


Questions of the impact of data on MPC performance in buildings, and the optimal data collection strategies to support model development, have been explored in the System Identification field \cite{zhan2021DataRequirementsPerformance,balali2023EnergyModellingControl,zhang2023InvestigationsMachineLearningbased,erfani2023LinkingDatasetQuality,zhan2022ImpactOccupantRelated,zhan2022ModelcentricDatacentricPractical} in the context of developing accurate system dynamics models for MPC, termed `control-oriented models'. Works have investigated the data requirements of different modelling approaches \cite{zhan2021DataRequirementsPerformance,balali2023EnergyModellingControl,zhang2023InvestigationsMachineLearningbased}, the impact of data resolution on model accuracy \cite{erfani2023LinkingDatasetQuality}, the impact of model prediction accuracy on operational performance \cite{zhan2022ImpactOccupantRelated}, and cost optimal data collection strategies to support model development \cite{zhan2022ModelcentricDatacentricPractical}.

\subsection{Research objectives \& novel contributions}


In the existing literature, the impact of data on the prediction accuracy of forecasting models for building energy management has been studied to a very limited extent. Additionally, there has been no study of the trade-off between {\color{\hlcolor}the quantity of data \& data features used for model training}, and the operational performance of MPC for battery scheduling in systems with distributed generation and storage. Understanding of this trade-off is necessary to properly prioritise expenditure on data collection for smart energy storage systems. This study aims to address this research gap by quantifying the impacts of data on both the prediction accuracy and operational performance of MPC using simple and high-complexity, state-of-the-art machine learning based prediction models. Simulation of a multi-building energy system with distributed solar generation and battery storage using historic building load measurement data is used as a case study. The main objectives of this study are:
\begin{itemize}
    \itemsep0em 
    \item Compare the performance of simple and state-of-the-art machine learning models with regards to prediction accuracy, model generalisation, and data efficiency;
    \item Investigate the trade-off between data and forecast accuracy for the following data efficiency measures: reuse of prediction models, reduction of training data {\color{\hlcolor}durations}, reduction of model data features, and online model training;
    \item Propose strategies for improving prediction performance when selecting models for reuse and selecting data to exclude when reducing training data {\color{\hlcolor}durations}; and
    \item Quantify the relationship between forecast accuracy and resulting operational performance of MPC.
\end{itemize}

The key contribution of this work is the combined study of the impact of data on both forecast accuracy and the resulting operational performance of MPC. This is important as it allows energy system designers to assess the trade-off between the cost of data for forecasting and the operational benefits it provides. The impact of aspects of data on forecast accuracy not yet studied in the context of building energy management are also investigated, specifically the reuse of prediction models, selection of model data features, and online model training. Two strategies for improving the efficacy of collected data for building load forecasting are proposed: a load profile similarity metric for selecting prediction models for reuse, and a change-point detection based methodology for screening training data to improve prediction performance whilst reducing model training time. Further, a long duration (10 year) historic building energy dataset is used to conduct these experiments, allowing performance to be evaluated on a multi-year scenario, providing more robust testing than existing studies.

The remainder of this work is structured as follows. Section \ref{sec:experiment} describes the simulation framework and data used to perform the experiments, as well as the models tested, and how their performance is evaluated. In Section \ref{sec:baseline-comparison}, the forecast accuracy of the prediction models trained without data {\color{\hlcolor}duration} limitations is evaluated to provide a baseline, and model performance is compared. Model generalisation for load prediction between buildings is then tested in Section \ref{sec:generalisation}, to assess whether model reuse is a viable strategy for reducing data collection requirements for new smart energy storage systems. A load profile similarity metric based on the Wasserstein distance between {\color{\hlcolor}functional Principal Component Analysis (fPCA)} coefficient distributions is proposed, and its efficacy as a criterion for selecting models for reuse is tested. Section \ref{sec:data-efficiency} studies the impact of various aspects of data on forecasting accuracy. The effect of the volume of data used for model training is studied in Section \ref{sec:data-efficiency-training-data} to support decision making on the quantity of data that should be collected for model development. A change-point detection based methodology for screening training data is proposed, and its ability to improve prediction accuracy whilst reducing {\color{\hlcolor}training data durations} is investigated. The selection of model data features and use of online model training are considered in Sections \ref{sec:data-efficiency-data-features} and \ref{sec:online-training} respectively. Section \ref{sec:control-sensitivity} contextualises the study of model forecast accuracy in the building energy system control task by quantifying the relationship between forecast accuracy and the resulting MPC operational performance for synthetic noisy forecasts. Finally, conclusions are drawn in Section \ref{sec:conclusions}.
\section{Experimental setup} \label{sec:experiment}


\subsection{Smart building control simulation framework}

\begin{figure}[h]
    \centering
    \includegraphics[height=5.5cm]{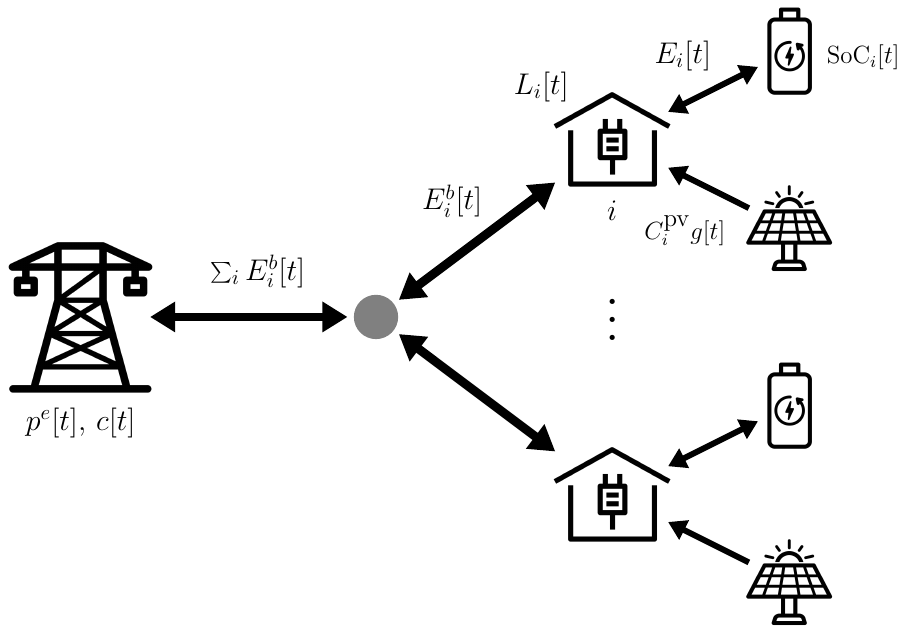}
    \caption{Energy flow schematic for test multi-building energy system. (Icon credits: \href{https://thenounproject.com/symbolon/}{Symbolon})}
    \label{fig:energy-system}
\end{figure}

To study the impacts of data on forecasting and MPC performance in the context of building energy systems, a {\color{\hlcolor}case study of an example multi-building building energy system was simulated. In this case study}, MPC is used to schedule battery storage operation in the multi-building energy system containing distributed storage and solar generation, as to reduce the electricity price, carbon emissions, and grid impact associated with meeting the electrical demand of the buildings. The CityLearn \cite{vazquez-canteli2019CityLearnV1OpenAI,vazquez-canteli2020CityLearnStandardizingResearch} building energy control framework is used to simulate the behaviour of the building energy system, and provide the required data to a Linear Program based MPC implementation. A schematic of the energy flows within the simulated multi-building energy system is provided in Fig. \ref{fig:energy-system}.

During simulations, at each time step, the prediction models use observation data to produce forecasts of the operational variables, which are passed to a linear predictive control model. The resulting linear optimisation problem is solved to determine the optimal control actions, which are then applied to the battery units in the CityLearn simulation. The combination of prediction models and linear predictive control model comprise the Linear MPC controller.


The Linear Program formulation used in the MPC scheme is described by Eq. \ref{eq:LP-example}, with Table \ref{tab:LP-params} providing descriptions of the parameters. At each time step the optimised control actions, ${E_i}^*[\tau{=}0]$, are taken. The optimisation objective is comprised of three weighted components, which correspond to the cost of grid electricity consumed by the buildings (assuming no net electricity metering), the embodied carbon emissions associated with the grid electricity, and the ramping of the overall grid electrical demand which represents the grid impact. The three components are normalised by the values they would have if no battery storage were present in the buildings, denoted by $\widetilde{O}^k$ for component $k$, lower bounded at 1. This clipping is performed to prevent ill-conditioning of the objective when the no-storage objective values are small.


The CityLearn simulations are configured so that the building energy system has linear dynamics, which is possible as a solar-battery system is studied and only electrical behaviour is considered. As the system parameters are known to the controller, the MPC has a perfect model of the true system dynamics. This perfect match between the simulator and controller dynamics models means no inaccuracies are introduced in system identification. The optimality guarantees of Linear Programming, and the use of the global operational objective as the MPC objective, with sufficiently large planning horizon $T$, mean there is negligible distortion of the operational performance from these factors. This allows the effect of operational condition forecast accuracy on MPC operational performance to be studied in isolation - i.e. to a good approximation, sub-optimality in operational performance in the simulation environment is caused solely by forecasting inaccuracies.

\begin{subequations} \label{eq:LP-example}
    \begin{align}
        \addtocounter{equation}{-1}
        \begin{split}
        \min & \qquad \frac{\gamma^p \,\, \mathlarger{\sum}_{\tau} \,\, p^e[t{+}\tau] \, \mathlarger{\sum_i} \left( \max \left[ 0,\, E_i^b[t{+}\tau] \right] \right)}{\raisebox{-1.2ex}{$\max \left[ 1, \widetilde{O}^p \right]$}} \,\, + \,\, \frac{\gamma^c \,\, \mathlarger{\sum}_{\tau} \,\, c[t{+}\tau] \, \mathlarger{\sum_i} \left( \max \left[ 0,\, E_i^b[t{+}\tau] \right] \right)}{\raisebox{-1.2ex}{$\max \left[ 1, \widetilde{O}^c \right]$}} \\[1ex]
        & \qquad\qquad\qquad\qquad + \, \frac{\gamma^r \,\, \mathlarger{\sum}_{\tau} \,\, \abs{\bigg( \mathlarger{\sum_i} \, E_i^b[t{+}\tau] \bigg) - \bigg( \mathlarger{\sum_i} \, E_i^b[t{+}\tau{-}1] \bigg)} }{\raisebox{-1.2ex}{$\max \left[ 1, \widetilde{O}^r \right]$}}
        \end{split} \label{eq:lp} \\[1ex]
        \text{over} & \qquad E_i[\tau] , \, \textrm{SoC}_i[\tau{+}1] \quad \forall \: i,\, \tau \tag*{} \\
        \text{subject to} & \qquad \textrm{SoC}_i[\tau{+}1] \leq \textrm{SoC}_i[\tau] + \min\left[ E_i[\tau] \sqrt{\eta_i},\, E_i[\tau]/\sqrt{\eta_i} \right] \label{eq:dynamics-constraint} \\
        & \qquad -P^{\textrm{max}}_i \Delta t \leq E_i[\tau] \leq P^{\textrm{max}}_i \Delta t \label{eq:power-constraint} \\
        & \qquad 0 \leq \textrm{SoC}_i[\tau{+}1] \leq C^s_i \label{eq:energy-constraint} \\
        & \qquad \textrm{SoC}_i[\tau{=}0] = \textrm{SoC}_i^t \label{eq:initial-conditions} \\
        & \qquad E^b_i[\tau] = L_i[\tau] - C^{\textrm{pv}}_i g[\tau] + E_i[\tau] \label{eq:aggregation-constraint} \\
        \text{for all} & \qquad i \in [0,B{-}1], \: \tau \in [0,T{-}1] \tag*{}
    \end{align}
\end{subequations}

\begin{table}[h]
    \centering
    \renewcommand{\arraystretch}{0.7}
    \begin{tabularx}{\linewidth}{ccX} \toprule \toprule
        Parameter & Units & \multicolumn{1}{>{\centering\arraybackslash}c}{Description} \\
        \midrule \midrule
        \multicolumn{3}{>{\centering\arraybackslash}l}{\small \quad Decision variables} \\
        $E_i[\tau]$ & kWh & Energy \textit{intake} to battery unit in building $i$ at time $\tau$ in planning horizon \\
        $\textrm{SoC}_i[\tau]$ & kWh & State-of-charge of battery unit in building $i$ at time $\tau$ in planning horizon \\
        \midrule
        \multicolumn{3}{>{\centering\arraybackslash}l}{\small \quad Data variables} \\
        $\Delta t$ & hrs & Time step of simulation data \\
        $C^s_i$ & kWh & Energy capacity of battery unit in building $i$ \\
        $C^{\textrm{pv}}_i$ & kWp & Peak power capacity of solar PV unit in building $i$ \\
        $\eta_i$ & -- & Round-trip efficiency of battery unit in building $i$ \\
        $P^{\textrm{max}}_i$ & kW & Power capacity of battery unit in building $i$ \\
        $\textrm{SoC}_i^t$ & kWh &  State-of-charge of battery unit in building $i$ at time $t$ \\
        $L_i[t]$ & kWh & Electrical demand of building $i$ at time $t$ \\
        $g[t]$ & kW/kWp & Normalised generation power from solar PV at time $t$ \\
        $p^e[t]$ & £/kWh & Grid electricity price at time $t$ \\
        $c[t]$ & kgCO$_2$/kWh & Carbon intensity of grid electricity at time $t$ \\
        $\gamma^p$ & -- & Fraction of objective from electricity cost component \\
        $\gamma^c$ & -- & Fraction of objective from CO$_2$ emissions component \\
        $\gamma^r$ & -- & Fraction of objective from grid ramping component \\
        \bottomrule \bottomrule
    \end{tabularx}
    \caption{Description of Linear Program model parameters.} \label{tab:LP-params}
\end{table}

\subsection{Forecasting task \& performance evaluation}


For the MPC scheme used, Eq. \ref{eq:LP-example}, forecasts of the following operational condition variables over the planning horizon $T$ are required at each time instance, $t$:
\begin{itemize}
    \item electrical demand for each building, $L_i[t]$
    \item normalised solar PV generation power, $g[t]$
    \item price of grid electricity, $p^e[t]$
    \item carbon intensity of grid electricity, $c[t]$
\end{itemize}

This study investigates the impacts of data on the forecasting of these 4 types of operational variables. The accuracies of these forecasts are quantified using two error metrics, the normalised Mean Absolute Error (nMAE), and the normalised Root Mean Squared Error (nRMSE), given by Eqns \ref{eq:nMAE} \& \ref{eq:nRMSE},

\vspace*{0.5cm}
\begin{minipage}{0.5\linewidth}
\begin{equation} \label{eq:nMAE}
    \frac{\frac{1}{N} \mathlarger{\sum_{t=0}^{N-1}} \, \frac{1}{T} \sum_{\tau=1}^{T} \abs{f^v_{t,\tau} - v_{t+\tau} } }{\frac{1}{N} \sum_{t=0}^{N-1} v_t}
\end{equation}
\end{minipage}%
\begin{minipage}{0.5\linewidth}
\begin{equation} \label{eq:nRMSE}
    \frac{\frac{1}{N} \mathlarger{\sum_{t=0}^{N-1}} \sqrt{ \frac{1}{T} \sum_{\tau=1}^{T} \left(f^v_{t,\tau} - v_{t+\tau} \right)^2 } }{\frac{1}{N} \sum_{t=0}^{N-1} v_t}
\end{equation}
\end{minipage}
\vspace*{0.5cm}

where $f^v_{t,\tau}$ is the forecast of variable $v$ at time $t$ for time instance $\tau$ in the planning horizon, and $v_{t+\tau}$ is the true value of the target variable at time instance $t+\tau$. These error metrics are the means of the standard MAE and RMSE errors over all forecasting horizons considered in the simulation, normalised by the mean level of the target variables to allow comparability between forecast accuracies.

The operational performance achieved by the MPC scheme using a given set of prediction models is quantified by evaluating the objective specified in Eq. \ref{eq:LP-example} with the simulated behaviour of the building energy system resulting from the use of the controller.

All experiments conducted in this study test the same multi-building energy system, with distributed energy assets as specified in \ref{app:system-spec}, and use a planning horizon of $T=48$hrs, justification of this value is provided in \ref{app:tau}, along with objective component weights, $(\gamma^p,\gamma^c,\gamma^r) = (0.45,0.45,0.1)$, in both the MPC scheme and for operational performance evaluation.

\subsection{Cambridge Estates electricity use dataset}

A dataset of historic building electricity usage measurements from a set of buildings across the Cambridge University Estates covering the period 2010 to 2019 \cite{langtry2024CambridgeUniversityEstates} is used {\color{\hlcolor}to provide a case study for the} multi-building energy system. The dataset consists of 10 years of historic electricity usage data from 30 buildings of varying use types in the University Estate, such as lecture blocks, offices, laboratories, and museums, alongside weather observations and grid electricity price and carbon intensity data. {\color{\hlcolor}The electricity usage measurements record the total electrical load of each building, which includes lighting, plug loads, and plant equipment electricity consumption. It is assumed that none of the buildings have heat pumps or AC units installed, meaning the electricity usage does not include any contributions from space heating or cooling.}

In all, the dataset contains the following variables: building electrical load data, weather data for Cambridge from the Met Office MIDAS dataset \cite{metoffice2022MIDASOpenUK} (temperature and relative humidity) and \href{https://www.renewables.ninja/}{renewables.ninja} reanalysis model \cite{pfenninger2016LongtermPatternsEuropean,staffell2016UsingBiascorrectedReanalysis} (direct and diffuse solar irradiance), dynamic electricity pricing tariff data from Energy Stats UK \cite{energystatsuk2023HistoricalPricingData}, grid electricity carbon intensity from the National Grid ESO Data Portal \cite{nationalgrideso2020HistoricGenerationMix}, and temporal information including hour, day, and month indices, as well as daylight savings status. All data is available at hourly resolution. Further detail on the data sourcing and processing is available in reference \cite{langtry2024CambridgeUniversityEstates}.

The 10 years of available data is initially split into training, validation, and test datasets covering the following periods: train (2010 to 2015), validate (2016 to 2017), test (2018 to 2019). For all experiments the test data is kept the same, however the periods of data used to train the prediction models are altered in Section \ref{sec:data-efficiency}. Of the 30 buildings available, 15 are selected\footnote{Building numbers: 0, 3, 9, 11, 12, 15, 16, 25, 26, 32, 38, 44, 45, 48, 49.} for use in the experiments, such that they cover a wide range of building scales and provide a good mix of similarity and dissimilarity with respect to their electricity loads.

\subsection{Brief description of the prediction models}


Data and computational requirements, which vary across predictions models, are important considerations for the deployment of MPC based controllers in practical building energy systems. This work investigates the performance of 6 machine learning based prediction models which span a range of model characteristics; 3 simple neural models, and 3 high-complexity, state-of-the-art models. A brief description of each model architecture follows. Technical specifications of the model implementations used are provided in \ref{app:models}.

\subsubsection{Simple neural models}

Recent literature, \cite{zeng2023AreTransformersEffective}, has shown that simple neural models using Direct Multi-Step forecasting (DMS), where all predictions over the forecast window are generated concurrently, can outperform complex, transformer-based models using traditional, Iterated Multi-Step forecasting (IMS), in which a single-step forecaster is applied iteratively to generate a multi-step forecast. For all three simple neural architectures investigated, DMS forecasting is used.

\paragraph{Linear neural network (Linear)}

A Multi-Layer Perceptron (MLP) model that maps the inputs directly to the output without an activation function (non-linearity).

\paragraph{Residual multi-layer perceptron (ResMLP)}

A Residual MLPSkip model (MLP model with skip-connections), comprised of a single hidden layer with 128 neurons.

\paragraph{Convolutional neural network (Conv)}

A Convolutional Neural Network (CNN) model that contains convolution layers followed by a linear layer. The architecture used comprises two layers with kernel sizes of 6 and 12, with five and one channels, respectively.

\subsubsection{State-of-the-art machine learning models}

\paragraph{Temporal Fusion Transformer}

The Temporal Fusion Transformer (TFT) model \cite{lim2021TemporalFusionTransformers}, developed by Google, is an attention-based architecture that enables the fusion of data from multiple input sources to inform predictions. The neural structure contains features which allow for the learning of multiple underlying relationships across temporal scales, and the attention mechanism allows for interpretation of the model predictions, i.e. which data the model is exploiting to produce its forecasts. The model uses categorical covariates of date-time information, as well as temperature information for predicting building loads.

\paragraph{Neural Hierarchical Interpolation for Time Series Forecasting}

Neural Hierarchical Interpolation for Time Series Forecasting (NHiTS) \cite{challu2023NHITSNeuralHierarchical} is an MLP model which learns a set of basis functions at different frequencies that describe the underlying patterns in the training data, and produces forecasts by using hierarchical interpolation to combine predictions from the basis functions in a computationally efficient manner. It uses categorical covariates of date-time information.

\paragraph{DeepAR}

DeepAR is a Recurrent Neural Network (RNN) based model developed by Amazon \cite{salinas2020DeepARProbabilisticForecasting}, which has been widely applied in a range of research areas. It is a probabilistic forecasting model, but for this study only the mean prediction is used. The model uses categorical covariates of date-time information.
\section{Results \& Discussion}



\subsection{Baseline prediction accuracy comparison} \label{sec:baseline-comparison}

The prediction accuracy of each forecasting model trained using 8 years of training data, {\color{\hlcolor}the maximum available,} was evaluated to provide a baseline for the model architectures in a setting without data limitations. For brevity, forecast accuracy results are discussed for the nRMSE metric only, however equivalent results were found with the nMAE metric. Figures \ref{fig:baseline-load-comparison} \& \ref{fig:baseline-PCS-comparison} show that the simple Linear model achieves similar or better prediction accuracy (lower nRMSE values) compared to the complex, state-of-the-art models across all prediction variables. The Conv and ResMLP models provide similar accuracy when predicting building electrical loads, however both have significantly worse accuracy for the electricity price and carbon intensity prediction variables. Complex models achieve slightly better accuracy for electricity price and solar generation predictions, at most 8.5\% and 4.2\% lower nRMSE than the Linear model, achieved by NHiTS and TFT respectively. However, for some buildings the complex models exhibit very poor forecast accuracy for load predictions. These instances of poor accuracy are found to correlate with a measure of the similarity between the train and test datasets for building load, called the `Wasserstein similarity metric', which is described in \ref{app:similarity-metric} and is used for the study of model reuse in the following section. Fig. \ref{fig:baseline-load-similarity-correlation} plots the relationship between prediction accuracy and data similarity for each model, and shows that complex models provide poor prediction accuracy when the training and test data are significantly different. Hence, simple neural models provide better prediction generalisation under changes in building load dynamics between the training and test data, which is highly advantageous for application to practical systems, as occupant driven load dynamics may change after system installation, e.g. due to a change of building use.

\newcommand{\baselineFiguresHeight}{4.3cm}
\begin{figure}[t]
\centering
\subfloat[Comparison of baseline load prediction accuracy of forecasting models.]{
    \includegraphics[height=\baselineFiguresHeight]{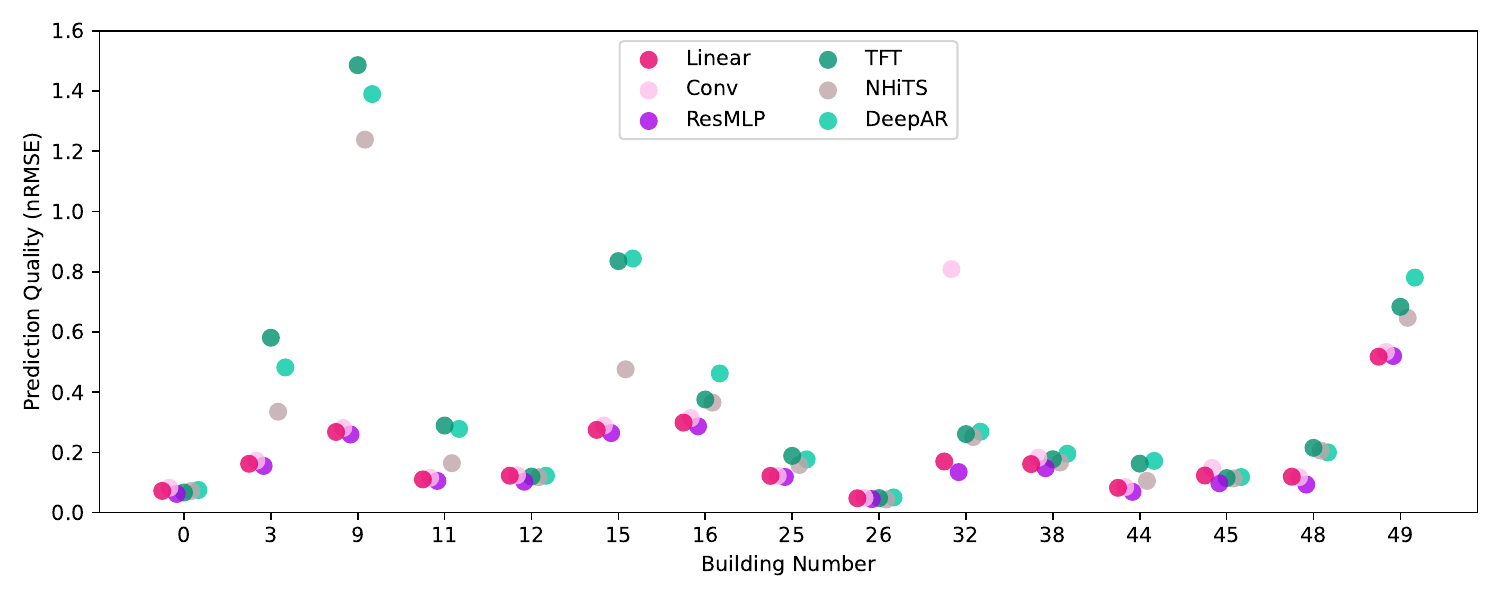}
    \label{fig:baseline-load-comparison}
}%
\hspace*{\fill}
\subfloat[Pricing, carbon, solar prediction comparison.]{
    \includegraphics[height=\baselineFiguresHeight]{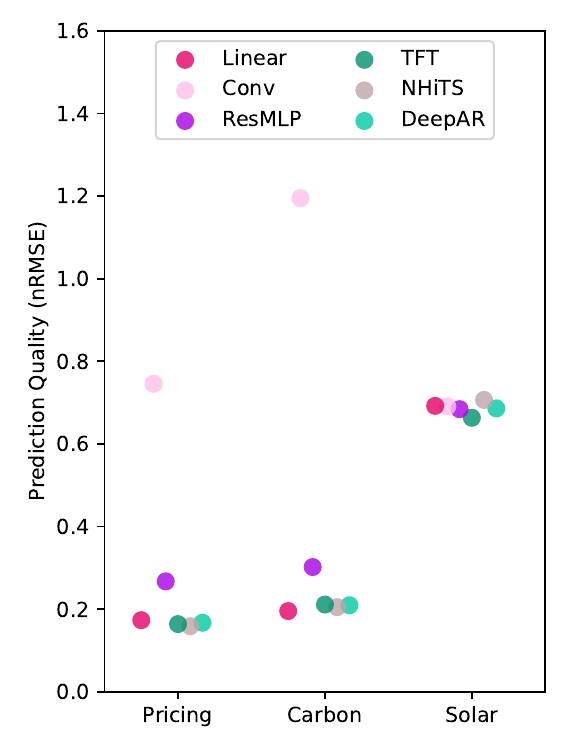}
    \label{fig:baseline-PCS-comparison}
}\hspace*{\fill}

\subfloat[Correlation between model prediction accuracy and train-test data\\similarity metric value (lower Wasserstein metric means more similar data).]{
    \includegraphics[height=\baselineFiguresHeight]{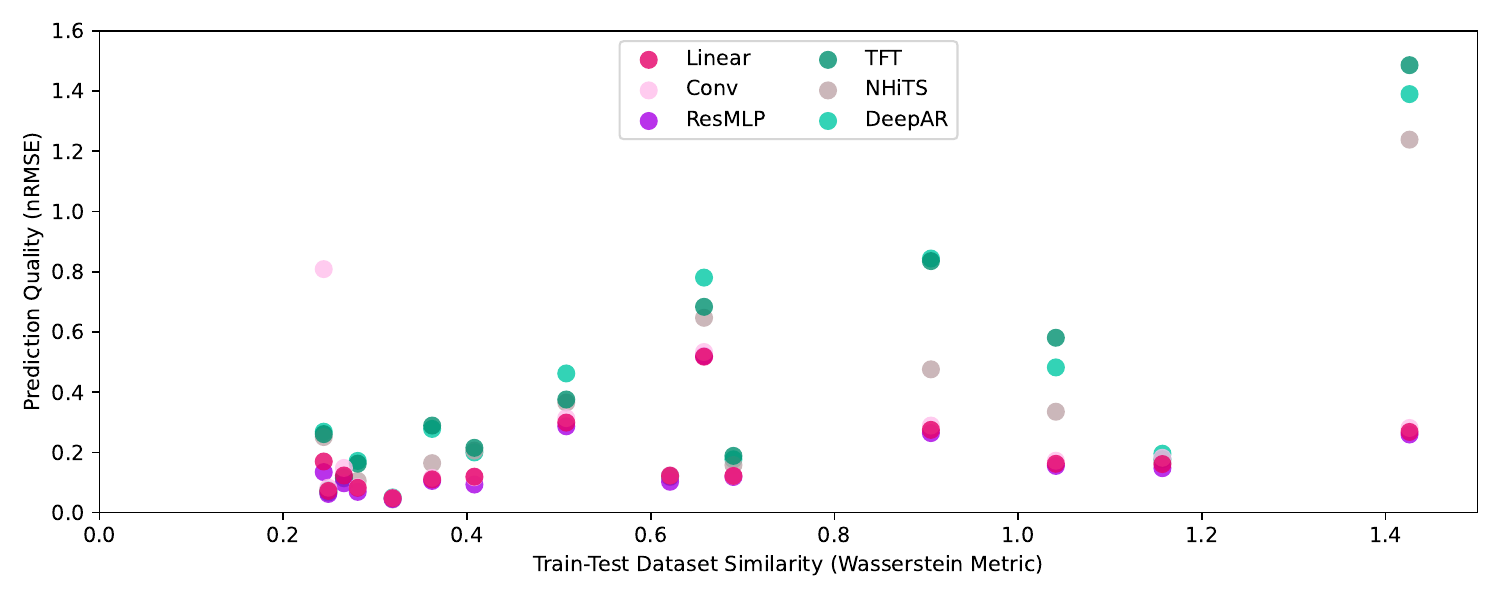}
    \label{fig:baseline-load-similarity-correlation}
}%
\hspace*{\fill}
\subfloat[Comparison of MPC operational performance using baseline forecasting models.]{
    \includegraphics[height=\baselineFiguresHeight]{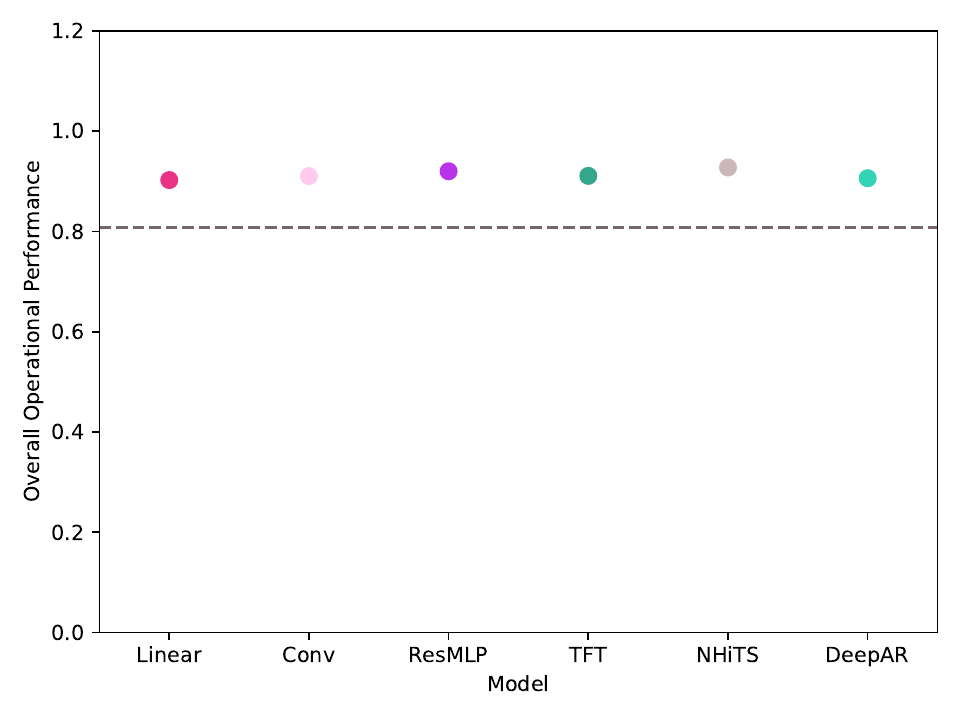}
    \label{fig:baseline-evaluate-comparison}
}\hspace*{\fill}

\caption{Comparison of baseline prediction accuracy and operational performance of forecasting models.}
\label{fig:baseline-comparison}
\end{figure}

{\color{\hlcolor}These results indicate that simple neural models have sufficient expressivity to well capture the underlying trends in building energy usage, but learn sufficiently simple relationships about the data as to avoid the problem of over-fitting experienced by the complex models, leading to better generalisation in time. This suggests that the majority of the trends in the load data are present within the 1 week input window used by the simple models, which agrees with the strong daily and weekly trends of typical building energy usage, caused by building occupancy patterns. It is likely that annual trends are well captured by the mean value over the input window, as these trends are slow relative to the prediction length. Additionally, the predictions required by MPC are relatively short in length, 48 hours in this case, whereas the complex models are found to provide more stable predictions and greater overall accuracy for longer duration forecasts \cite{challu2023NHITSNeuralHierarchical,zhang2022TemporalFusionTransformer}.}

Overall, simple neural models are able to provide analogous prediction accuracy to complex models across all prediction variables, and do so with substantially lower computational requirements. Table \ref{tab:comp-times} provides the training time and inference time (time to generate forecasts) of the baseline models, and shows that simple models require roughly 500x less computation for inference. Hence, in practical systems the use of simple models would enable shorter prediction intervals, leading to higher frequency control, and allow for the use of lower cost compute hardware. {\color{\hlcolor}Further, due to their high computational cost, the complex models were only able to use a 72 hour input window \cite{lopezsantos2022ApplicationTemporalFusion}. As a result, they have less information available to inform their predictions, likely limiting the accuracy they could achieve. Therefore, the computational efficiency of simple neural models enabling the use of longer input windows provides an additional advantage.}

\begin{table}[h]
    \centering
    \renewcommand{\arraystretch}{0.7}
    \begin{tabularx}{\linewidth}{c|*{6}{X}} \toprule \toprule
        Model & Linear & Conv & ResMLP & TFT & NHiTS & DeepAR \\ \midrule
        Training time (hrs) & 12.3 & 26.8 & 13.4 & 8.9 & 15.2 & 9.3 \\
        Prediction inference time (s) & 29 & 73 & 146 & 15,222 & 15,289 & 74,991 \\ \bottomrule\bottomrule
    \end{tabularx}
    \caption{Computation times for baseline models, trained on 8 years of data, and predicting for simulations of 2 years duration.}
    \label{tab:comp-times}
\end{table}

The similar prediction accuracies of the tested models are found to lead to similar operational performance when used in the MPC scheme, see Fig. \ref{fig:baseline-evaluate-comparison}, where the dashed line indicates the bound on operational performance achieved by an MPC scheme with perfect forecasts. The use of the MPC controllers with the specified battery systems leads to an average 8.7\% improvement in operational performance for the multi-building energy system, with a range of 7.3\% to 9.8\%.

\subsection{Model generalisation} \label{sec:generalisation}


When a solar-battery system using MPC is installed, high-resolution historic load metering data (e.g. from smart meters) may not be available for the building. In this case, the project must either be delayed to allow time for data collection, incurring a significant cost, or a prediction model trained on load data from another building must be used, potentially incurring an operational performance penalty due to worse prediction accuracy. Model reuse greatly reduces data collection requirements and the associated costs, however its appropriacy depends on the ability of load prediction models to generalise between buildings, and the trade-off between data cost savings and increased operational cost due to lower forecast accuracies.

\begin{figure}[t]
    \centering
    \begin{minipage}{.475\textwidth}
        \centering
        \includegraphics[height=5.75cm]{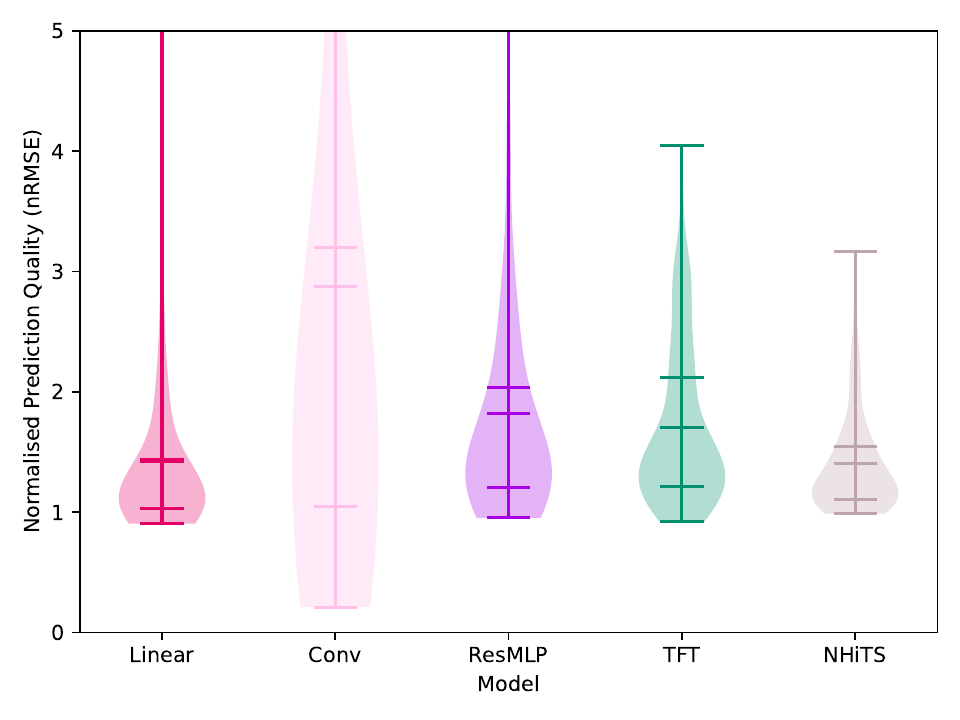}
        \caption{Comparison of forecasting model generalisation. Violin plot with quartiles indicated by horizontal lines.}
        \label{fig:generalisation-violin}
    \end{minipage}%
    \hfill
    \begin{minipage}{.475\textwidth}
        \centering
        \includegraphics[height=5.75cm]{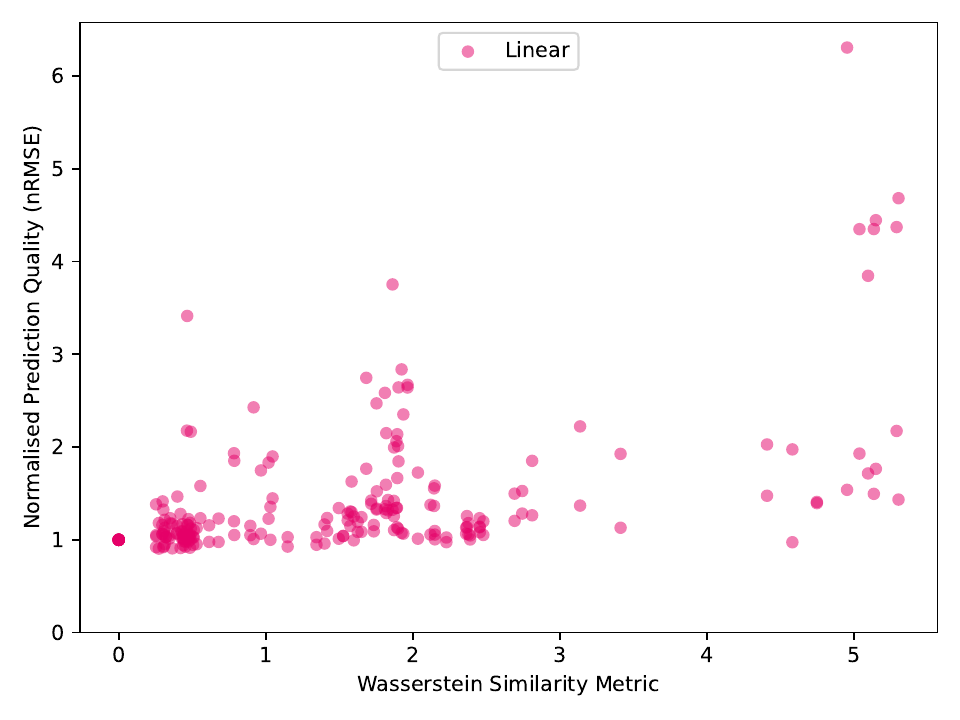}
        \caption{Correlation between Linear model generalisation and training dataset similarity metric values.}
    \label{fig:generalisation-linear-Wass-corr}
    \end{minipage}%
\end{figure}

The generalisation of prediction models between buildings is tested by using the baseline models trained on data from each of the 15 buildings to forecast electrical load for every other building. Fig. \ref{fig:generalisation-violin} shows the distribution of relative forecast accuracies achieved by each model over all buildings (DeepAR is excluded due to excessive computational costs), where the y-axis plots the nRMSE forecast accuracy of the tested model (potentially trained on a different building) normalised by the forecast accuracy achieved by the model trained on data from the target building. The Linear model provides similar generalisation performance to NHiTS, and is substantially better than all other models. {\color{\hlcolor}It is proposed that the Linear model, being mechanistically equivalent to linear regression, achieves good generalisation as it learns relatively simple relationships about the load data, which are consistent between buildings. Whereas NHiTS achieves good generalisation due to its behaviour of generating smooth forecasts \cite{challu2023NHITSNeuralHierarchical}, making it less susceptible to producing erratic and inaccurate predictions with unseen input data. As before, TFT suffers from over-fitting, leading to relatively poor generalisation. Whilst the Conv model was able to provide good generalisation in time where the dataset similarity was close, indicated by small Wasserstein metric values in Fig. \ref{fig:baseline-load-similarity-correlation}, between buildings the load data is much less similar, and the relationships learned by the Conv model are no longer valid, leading to poor prediction accuracy. It is suggested this is due to the features learned by the pooling process no longer being pertinent for the new building.}

If a pre-trained Linear model is selected for reuse at random, the average prediction accuracy is substantially worse than a model trained on data collected from the target building (45\% higher nRMSE, corresponding to the mean of the violin plot). However, in a real scenario where an ensemble of pre-trained models is available, the smart energy storage system designer could reuse a model from a building that is expected to be similar to the target building (e.g. by size, usage type, envelope characteristics, load dynamics, etc.), to maximise the likelihood of good generalisation. A load profile similarity metric is additionally proposed as a way off assessing the similarity of building load dynamics for the purposes of model reuse. This similarity metric is based on the similarity of the distributions of underlying mode shapes within each load profile, and is referred to as the `Wasserstein similarity metric' as it uses the Wasserstein distance between mode shape coefficient distributions. \ref{app:similarity-metric} describes the calculation of similarity metric values. The correlation between generalisation of the Linear model and the Wasserstein similarity metric values between the training datasets of the reused model building and target building is shown in Fig. \ref{fig:generalisation-linear-Wass-corr}. The cluster of points in the bottom left indicates that buildings with similar load profiles (low Wasserstein metric values) can provide models that achieve good generalisation. When selecting models for reuse by minimising the Wasserstein metric, the average relative prediction accuracy improves to an 11\% increase in nRMSE, with a range of -7.9\% to +52.6\%, showing that some reused models outperform those trained on the true building training dataset.

These results show that when developing a load forecasting model for a new building system where little or no data is available, the reuse of forecasting models trained on existing buildings provides a promising\footnote{Further work is required to determine the feasibility of identifying similar buildings from the model ensemble in practice. Whilst a small sample of load data from the target building could be used to approximate the Wasserstein similarity metric, the quality of this approximation must be weighed against the forecast accuracy that could be achieved by a model trained using that data. Though, additional information on the building, such as usage type, could be used to guide model selection for reuse.} alternative to the collection of data and training of a building specific model, and the associated costs. However, reusing prediction models increases uncertainty in the operational performance the MPC system will achieve, as models trained on load data from buildings similar to the target building do not always provide good prediction accuracy, as shown by the wide range of reused model accuracies.


\subsection{Data efficiency} \label{sec:data-efficiency}


Data required for developing forecasting models incurs cost from both its acquisition and the computation required to exploit it. However, {\color{\hlcolor}using longer durations of training data or additional data variables} can increase forecast accuracy, resulting in lower operational cost of the energy system. Therefore, any measures to reduce {\color{\hlcolor}the quantity of data used} must be traded-off against the corresponding decrease in prediction accuracy. This section explores a series of aspects of data efficiency and their effect on forecast accuracy. Subsequently, section \ref{sec:control-sensitivity} investigates the impact of forecast accuracy on operational performance of the MPC.

Testing data efficiency requires several versions of a prediction model to be trained. Due to computational limitations, for some data efficiency tests only the Linear prediction model is studied, as the complex models are too computationally expensive, and results from previous sections show the Linear model is the best performing simple model.

\subsubsection{Volume of training data} \label{sec:data-efficiency-training-data}

\begin{figure}[t]
    \centering
    \begin{minipage}{.475\textwidth}
        \centering
        \includegraphics[height=5.75cm]{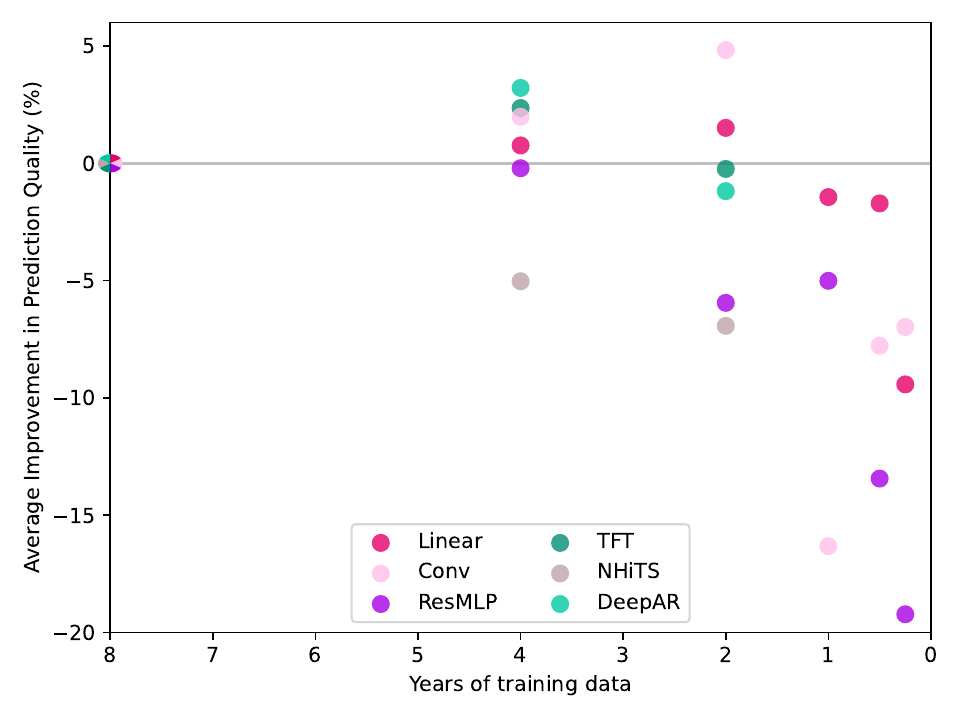}
        \caption{Average improvement in prediction accuracy over all prediction variables with years of data used for training.}
        \label{fig:data-eff-n-training-years}
    \end{minipage}%
    \hfill
    \begin{minipage}{.475\textwidth}
        \centering
        \includegraphics[height=5.75cm]{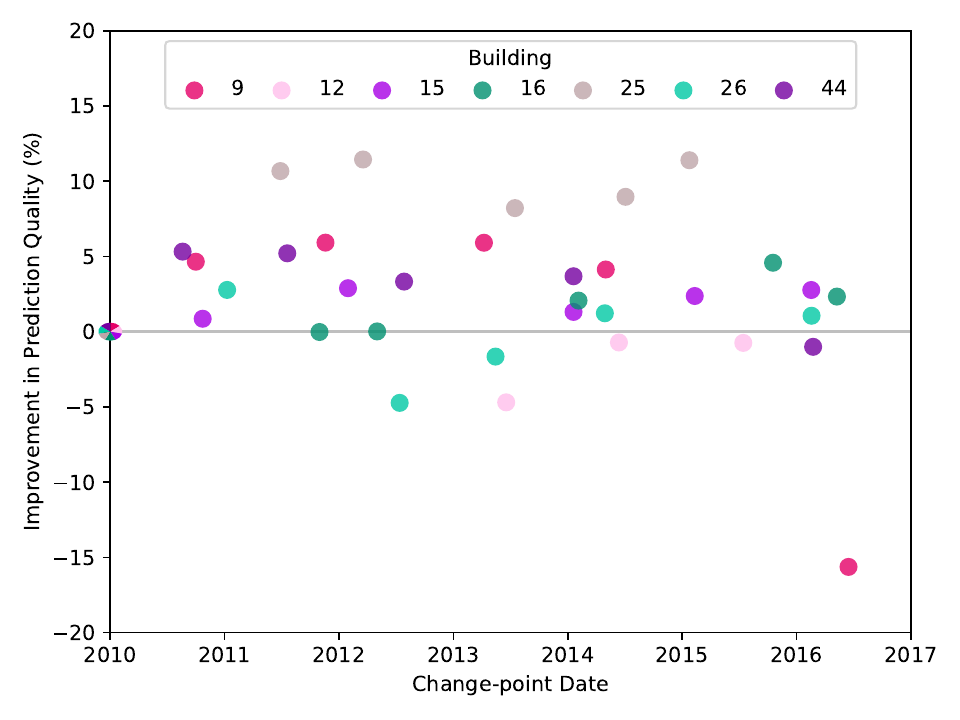}
        \caption{Improvement in prediction accuracy of models trained using data from change-points onward.}
    \label{fig:data-eff-changepoints}
    \end{minipage}%
\end{figure}

Using greater volumes of training data, longer durations of historic measurements of buildings, can improve model prediction accuracy by avoiding over-fitting, improving temporal generalisability, but comes at a roughly linear cost. The trade-off between the {\color{\hlcolor}length} of training data used, the combined duration of the train and validate datasets, and forecast accuracy is investigated by training prediction models on subsets of the building energy dataset of varying durations from 8 years, {\color{\hlcolor}the maximum duration available, to 3 months (one season), the shortest duration tested in \cite{choi2023PerformanceEvaluationDeep}}. For the complex models, at least 2 years of training data was required due to the use of temporal covariates. Fig. \ref{fig:data-eff-n-training-years} plots the proportional improvement in prediction accuracy of each model (negative improvement shows worsening model accuracy) compared to baseline when training using different data volumes. Across a broad range of model architectures, reducing training data {\color{\hlcolor}length} down to 2 years has a limited impact on prediction accuracy, in some cases improving prediction accuracy. Trends differ between prediction variables, however for the Linear model, in the majority of cases there is a small prediction accuracy penalty between 8 and 1 years of training data, and then a rapid worsening of forecasting accuracy below {\color{\hlcolor}6 months}.
This indicates that when making data collection decisions to support prediction model development for building electrical energy systems, at most 2 years of measurement data should be gathered. {\color{\hlcolor}However, at least two seasons of data (6 months) are required to prevent model over-fitting and learn trends which generalise across seasons.}

\paragraph{Screening training data using change-points} \label{sec:data-efficiency-change-point}


Once building data has been gathered, there remains a question as to whether all of the available data should be used for training, or whether some data should be excluded to improve model prediction. For instance, if data that is non-representative of the present building load dynamics can be excluded, prediction accuracy can be improved alongside data efficiency.
Change-point analysis can be used to screen training data for this purpose, by detecting changes in building load dynamics and excluding data preceding the change from training. A change-point analysis using the BEAST algorithm \cite{zhao2019DetectingChangepointTrend} was performed on the building load dataset, described in \ref{app:change-points}. Points where changes in load trend were detected were used to screen the training data for 7 selected buildings. In order to allow a better analysis of the detected change-points, the training dataset was increased to 7 years (2010-2016), leaving 1 year of validation data (2017). Linear prediction models were trained using sections of the training data starting at each of the detected change-points and running to the end of 2016.

Fig. \ref{fig:data-eff-changepoints} shows the relative improvements in prediction accuracy of models trained on data screened using change-points, compared to the case where the full 7 years of available data is used, where the x-axis position indicates the timing of the change-point. For most buildings, screening the training data using change-points improves prediction accuracy, indicating that non-representative data is being removed from the training dataset. This suggests that ex-post training data screening can be used to improve model accuracy whilst reducing {\color{\hlcolor}the quantity of training data used}, and so computational cost. These results correspond with the findings of \cite{choi2023PerformanceEvaluationDeep}, which shows that additional training data only improves prediction accuracy if it is sufficiently similar to the test data, and can reduce accuracy when insufficiently similar. However, the results of this study demonstrate an additional consideration, which is that prediction accuracy worsens substantially when insufficient training data volume is used. This suggests a trade-off between having enough data to avoid over-fitting, and removing non-representative data which causes over-generalisation of the model. Figures \ref{fig:data-eff-n-training-years} \& \ref{fig:data-eff-changepoints} indicate that at least 1 year of training data is required to achieve good prediction accuracy, after which only sufficiently similar data should be included in the training dataset. The reason for the difference in results compared to \cite{choi2023PerformanceEvaluationDeep} is likely due to this study testing model prediction accuracy on 2 years of data, compared to 0.1 years in \cite{choi2023PerformanceEvaluationDeep}. As a result, models in this study must capture the seasonality of building load dynamics, hence requiring at least 1 year of training data to achieve good prediction accuracy.

The data screening method considered in this study is relatively simple, using change-points to exclude data by classifying it as `non-representative' of current behaviour. More advance techniques could be created by quantifying the similarity between the sections of data identified using the change-points and the validation data, e.g. using the Wasserstein similarity metric, and selecting sections of data to use in the training dataset using both the probability of a true change having occurred and the data similarity metrics.

{\color{\hlcolor}Change-point analysis can also be used online to detect real time changes in building load dynamics and provide indication of when the prediction models should be updated (i.e. retrained using recent data), due to the training data of the current model no longer being representative of the building behaviour, causing reduced prediction accuracy. This is particularly pertinent in the context of climate change, which is expected to significantly impact the energy usage behaviour of buildings, for instance reducing peak heating loads due to higher winter temperatures, and increasing summer electrical loads to provide cooling during prolonged heat waves.}

\subsubsection{Data features} \label{sec:data-efficiency-data-features}

\begin{figure}
    \centering
    \includegraphics[width=0.5\linewidth]{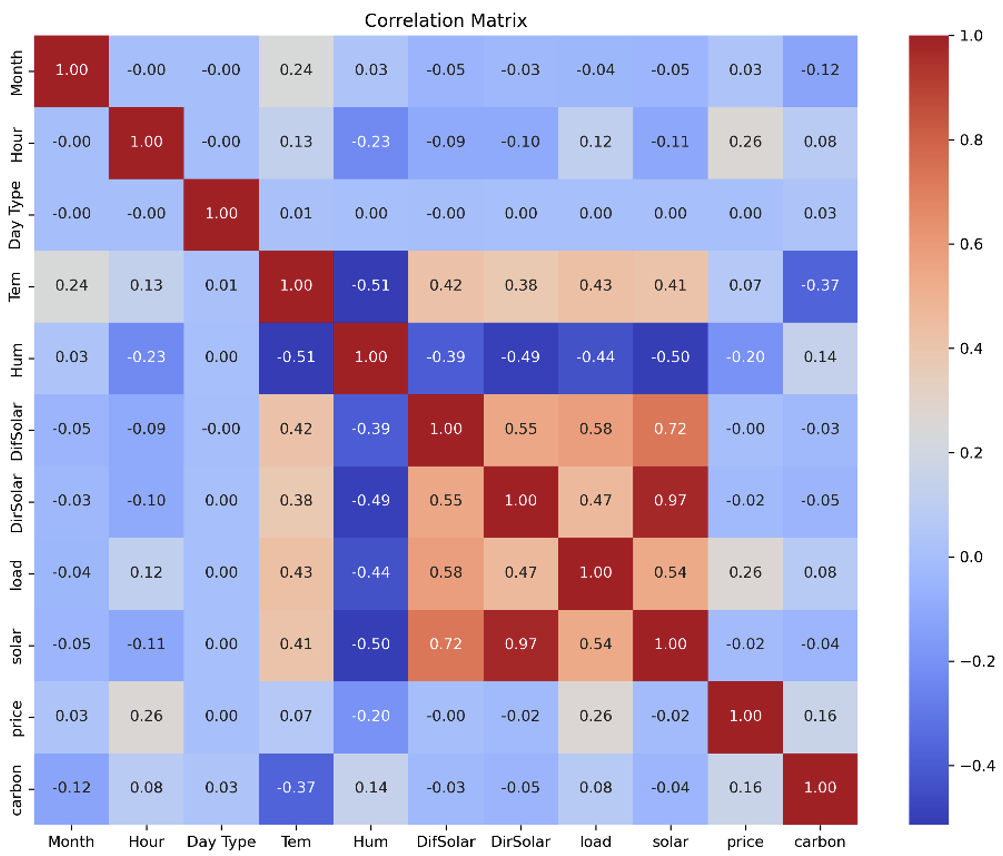}
    \caption{Pearson correlation between data variables for Building 0.}
    \label{fig:data-eff-variable-correlations}
\end{figure}
\begin{figure}
    \centering
    \includegraphics[width=0.9\linewidth]{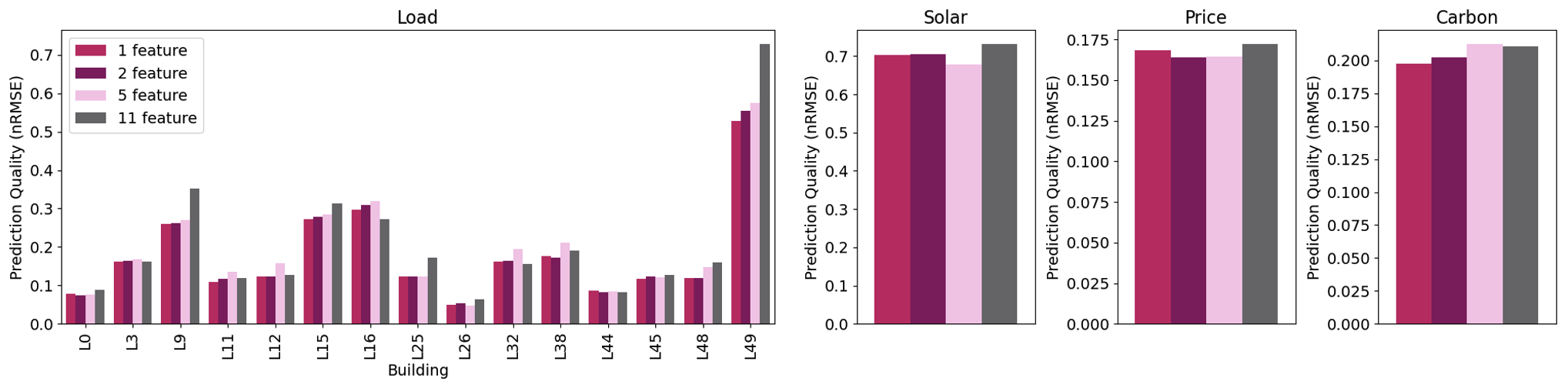}
    \caption{Variation of prediction accuracy with number of data features included in Linear model.}
    \label{fig:data-eff-n-variables}
\end{figure}

The variables that can be monitored in building energy systems are often well correlated, for example ambient temperature and electrical load. As a result, covariates can be used by prediction models to attempt to identify underlying links between the variables and improve forecast accuracy. However, the use of additional data variables (features) incurs a cost from both the collection and exploitation of the extra data, for instance the purchase of proprietary weather forecasts. The impact of feature selection on forecast accuracy is investigated by comparing the prediction accuracy of models trained using varying numbers of data features. The Pearson correlation coefficient between the 11 data features available in the building energy dataset are shown for the case of Building 0 in Fig. \ref{fig:data-eff-variable-correlations}. Linear prediction models were trained using the $n$ most correlated variables/features for each of the electrical load, solar generation, electricity price, and carbon intensity target variables. {\color{\hlcolor}Feature selection was performed separately for each case by ranking Pearson correlation coefficients between the target variable and all other available covariates}. Fig. \ref{fig:data-eff-n-variables} shows the impact of {\color{\hlcolor}the number of data features used by the models} on forecast accuracy. For most prediction variables, the inclusion of additional data features worsens forecast accuracy. Several factors could contribute to this behaviour:
\begin{itemize}
    \item Over-fitting \cite{hawkins2004ProblemOverfitting} : The incorporation of an excessive number of features may lead the model to over-fit the training data, worsening its predictions for unseen data.
    \item Multi-collinearity \cite{farrar1967MulticollinearityRegressionAnalysis} : The presence of highly correlated features can destabilize the model, resulting in worse prediction accuracy.
    \item Curse of Dimensionality \cite{verleysen2005CurseDimensionalityData} : An increase in the number of features increases the dimensionality of the data, which can lead to data sparsity and degradation of prediction accuracy.
\end{itemize}

These results show that testing should be performed before decisions are made regarding the collection of covariate data to ensure that, for the type of system in question, its use will firstly improve prediction accuracy, and secondly that said accuracy improvements warrant the cost of the data. {\color{\hlcolor}This study considers a simple, correlation based feature selection method, however more advanced, optimization based techniques \cite{gonzalez-vidal2019MethodologyEnergyMultivariate,kim2020ElectricityLoadForecasting} could be used to determine the set of available features which provides the optimal trade-off between data cost and model prediction accuracy.}

\begin{figure}
    \centering
    \includegraphics[width=0.9\linewidth]{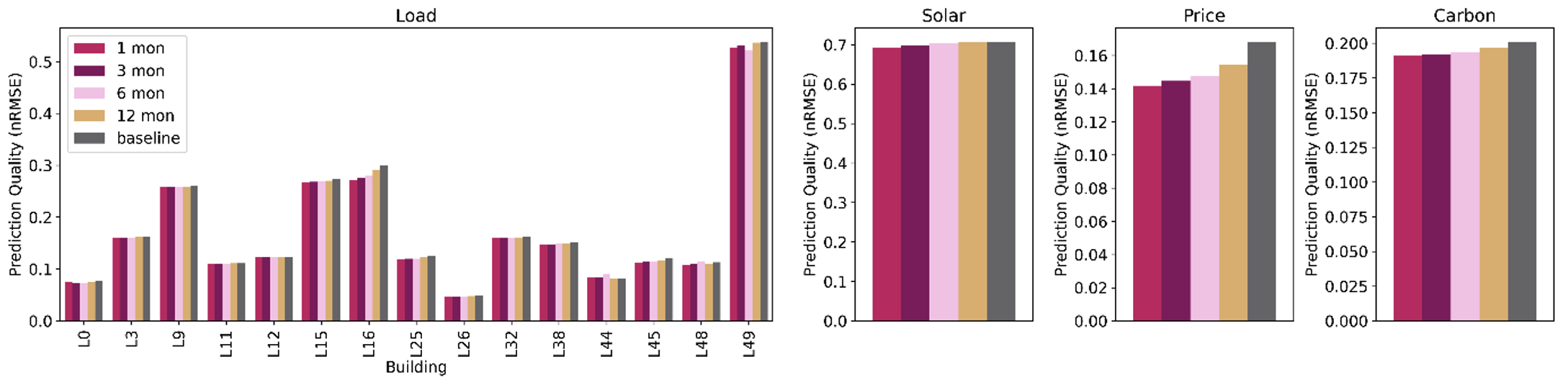}
    \caption{Variation of prediction accuracy with online update frequency.}
    \label{fig:data-eff-update-freq}
\end{figure}

\subsubsection{Online training} \label{sec:online-training}


During building operation monitoring systems continuously collect operational data. The characteristics of building behaviour can change during operation due to external factors such as weather \& climate, occupancy, and equipment degradation \& maintenance. Continuous online training updates the predictive models using the collected monitoring data, improving the model's ability to adapt to dynamic changes in building behaviour. A version of the Linear prediction model with online training is implemented to investigate the impact of online training frequency on model accuracy. {\color{\hlcolor}The model is updated online by tuning the model parameters (retraining) using a sliding window of training data with length equal to the update frequency.}
The prediction accuracy of models updated at different frequencies is plotted in Fig. \ref{fig:data-eff-update-freq}. Higher update frequencies led to increased prediction accuracy across all prediction variables. For example, prediction accuracy for grid electricity price improved by 15.7\%, 14.0\%, 12.1\%, and 8.1\% when updated monthly, quarterly, semi-annually, and annually, respectively, compared to the baseline model which is not trained online. {\color{\hlcolor}Online training allows the prediction model to adapt to changes in the underlying trends which occur over time, as it can learn the trends in recent data. This improves prediction accuracy as the model does not need to generalise in time, as it is trained (updated) on data that is representative of the current prediction horizon.}
Therefore, it is expected that the additional computational hardware and system complexity required for online trained prediction models will be worthwhile in practical systems.

\subsection{Sensitivity of Model Predictive Control to forecast accuracy} \label{sec:control-sensitivity}


Whether investments in additional data to improve forecast accuracy provide net benefit to the operation of a smart energy storage system depends on the resulting improvements in operational performance achieved by the MPC controller. The determination of optimal data {\color{\hlcolor}collection} strategies therefore requires quantification of the relationship between forecast accuracy and MPC operational performance. To investigate this in a controlled setting, the operational performance achieved by the tested MPC scheme is evaluated with the use of synthetic forecasts. The synthetic forecasts are produced by adding a Gaussian random walk noise component to the ground-truth values, as described in Eq. \ref{eq:GRWN},

\begin{equation} \label{eq:GRWN}
    f_{\mathsmaller{\textrm{GRW}}}^v[t,\tau] = v_{t+\tau} + \sum_{j=1}^{\tau} w_j^{\sigma} \qquad \textrm{s.t.} \:\: w_j^{\sigma} \:\: \textrm{i.i.d.} \:\: \mathcal{N}\!\left( \mu{=}0, \sigma^2 \right)
\end{equation}

where $v_{t+\tau}$ is the ground-truth value of variable $v$ at instance $\tau$ in the planning horizon of the forecast created at time $t$, and the noise level $\sigma$ can be selected.

\begin{figure}
    \centering
    \subfloat[Variation of operational performance components with amplitude of noise on all prediction variables.]{
        \includegraphics[height=5.75cm]{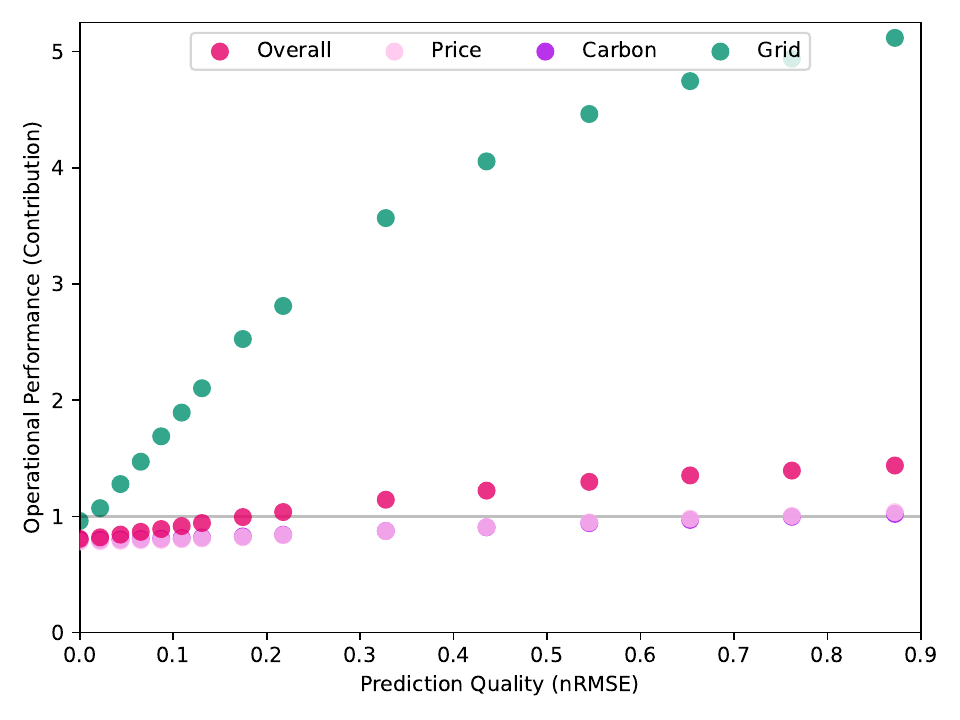}
        \label{fig:control-sens-obj-vs-noise}
    }%
    \hfill
    \subfloat[Variation of overall operational performance with amplitude of noise on each type of prediction variable.]{
        \includegraphics[height=5.75cm]{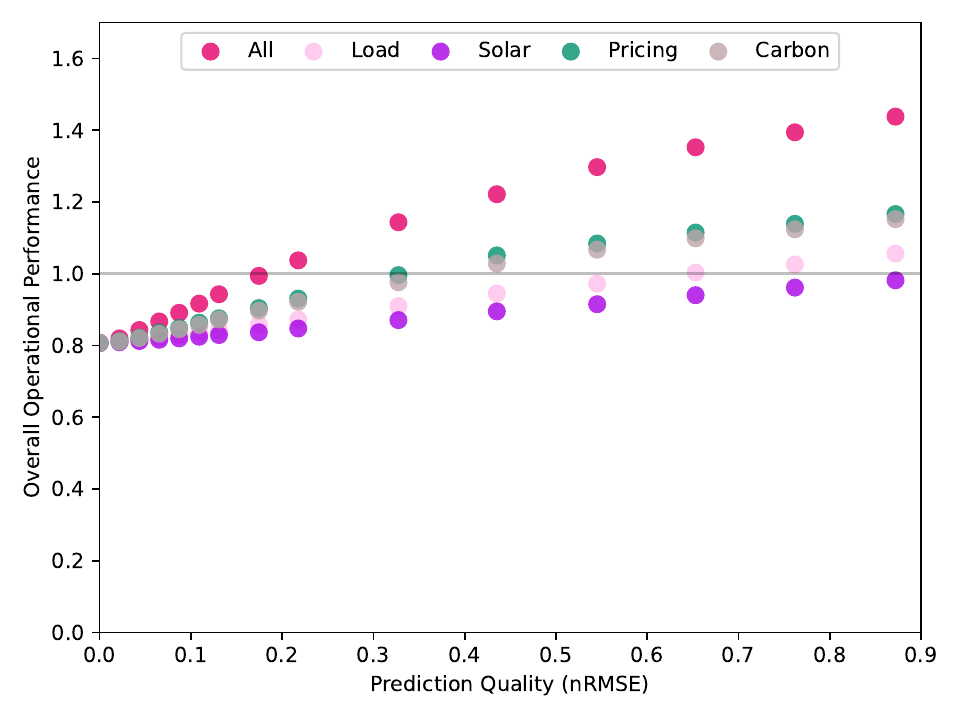}
        \label{fig:control-sens-control-vs-noise-type}
    }
    \caption{Sensitivity of operational performance to synthetic forecast noise simulating prediction inaccuracy.} \label{fig:control-sens}
\end{figure}

The MPC scheme using these synthetic forecasts for all prediction variables is tested at varying noise levels. Fig. \ref{fig:control-sens-obj-vs-noise} shows the resulting operational performance of MPC, as well as the three components it is comprised of (electricity price, carbon emissions, and grid impact), as the prediction accuracy of the synthetic forecasts varies. Fig. \ref{fig:control-sens-control-vs-noise-type} shows the variation in overall operational performance of the MPC when synthetic forecasts are used for each type of prediction variable in turn, with perfect forecasts used for all other variables. The horizontal line at 1 indicates the performance of the building energy system without battery control, which is the point at which the MPC controller becomes redundant.

The results show that the tested MPC scheme is most sensitive to the forecast accuracy of the grid electricity price and carbon intensity variables, suggesting that the most resource and expense should be invested in producing accurate forecasts of the grid conditions. Additionally, whilst the prediction models tested in this study performed worst when forecasting solar generation, see Fig. \ref{fig:baseline-PCS-comparison}, it may not be worth expending additional resource to improve these forecasts as the MPC scheme is least sensitive to this variable. Though the grid component of operational performance is found to be by far the most sensitive to forecast accuracy, this is likely due to the synthetic forecast model used\footnote{In the synthetic forecast model used, the Gaussian random walk noise component is regenerated at each time instance. This leads to the generation of forecasts which are correct on average, but can cause the MPC controller to take opposing energy management strategies in successive time steps, leading to large magnitudes of energy flows from the battery to correct the strategy. Further work is required to create a noise model for synthetic forecasts that is calibrated to the operational performance vs forecast accuracy trade-off of practical prediction models. With such a noise model, the experimental method demonstrated could be used to quantify the economic advantage of forecast accuracy improvements to support decision making regarding data {\color{\hlcolor}collection} for the development of forecasting models in practical smart energy storage systems.}, and is not considered to be reflective of the behaviour of real prediction models.
\newpage
\section{Conclusions} \label{sec:conclusions}


This study investigated the impacts of data on the prediction accuracy of forecasting models for building operational conditions, and the resulting operational performance of a Model Predictive Control (MPC) scheme in a multi-building energy system with distributed generation and storage. Experiments were conducted using a large-scale dataset of electrical load measurements from buildings in the Cambridge University Estates.

A simple linear multi-layer perceptron model (Linear) using DMS prediction was found to achieve equivalent forecast accuracy to high-complexity, state-of-the-art machine learning models in a setting without data limitations, but has substantial advantages regarding data efficiency and generalisation performance. Therefore, this simple neural model is preferable for use in practical MPC systems due to its lower data and computational requirements, and better performance on new load dynamics.

Using more than 2 years of hourly resolved training data did not provide significant improvements in prediction accuracy for most of the models tested, indicating that the collection of monitoring data for longer durations is unnecessary for the development of performant MPC schemes. Further, screening training data using change-point analysis to remove low similarity data was able to simultaneously improve data efficiency and prediction accuracy, provided at least 1 year of training data was kept. This could also be achieved through the removal of redundant data features from models.

The reuse of Linear models for load prediction between buildings was shown to be an effective way of reducing data collection requirements. When selecting models for reuse using a proposed load profile similarity metric based on the Wasserstein distance between fPCA coefficient distributions, model reuse led to an average 11\% increase in prediction error. However in comparison, models trained using only 3 months of building specific data provided forecasts with an average 9.9\% increase in error compared to the baseline. Additionally, online training of prediction models was shown to be highly beneficial, with higher update frequencies providing greater forecast accuracy improvements, of up to 15.7\% for the case of monthly updating. Hence, monitoring data should be used to update reused prediction models in situ to tailor them to the building system, ultimately replacing the reused model. The results suggest that, by exploiting existing building energy datasets to pre-train models, in many cases sufficient forecast accuracy can be achieved without the collection of any building load data prior to the installation of the system.

The relationship between forecast accuracy and operational performance of the MPC controller was investigated for synthetic forecasts, which showed the MPC scheme is most sensitive to grid electricity price and carbon intensity prediction accuracy. This analysis methodology would allow decision makers to determine whether expenditure on data collection to produce forecasting models is economic for a practical building energy system, i.e. whether the costs of data are outweighed by the improvements in operational performance provided.


\section*{CRediT authorship contribution statement}

\textbf{Max Langtry}: Conceptualization, Software, Methodology, Investigation, Writing - Original Draft, Project administration
\textbf{Vijja Wichitwechkarn}: Methodology
\textbf{Rebecca Ward}: Investigation
\textbf{Chaoqun Zhuang}: Methodology, Investigation
\textbf{Monika J. Kreitmair}: Investigation
\textbf{Nikolas Makasis}: Methodology
\textbf{Zack Xuereb Conti}: Methodology
\textbf{Ruchi Choudhary}: Conceptualization, Supervision, Writing - Review \& Editing

\section*{Declaration of competing interests}

The authors declare that they have no known competing financial interests or personal relationships that could have appeared to influence the work reported in this paper.

\section*{Data availability}

All data used in this study and the code used to train and test the forecasting models is available at \url{https://github.com/EECi/Annex_37}.

\section*{Acknowledgements}

Max Langtry is supported by the Engineering and Physical Sciences Research Council, through the CDT in Future Infrastructure and Built Environment: Resilience in a Changing World, Grant [EP/S02302X/1].

Vijja Wichitwechkarn is supported by the Engineering and Physical Sciences Research Council, through the CDT in Agri-Food Robotics: AgriFoRwArdS, Grant [EP/S023917/1].

Chaoqun Zhuang, Rebecca Ward and Zack Xuereb Conti are supported by the Ecosystem Leadership Award under the EPSRC Grant [EP/X03870X/1], and The Alan Turing Institute, particularly the Turing Research Fellowship scheme under that grant.

Monika J. Kreitmair and Nikolas Makasis are supported by the Surrey Future Fellowship program at the University of Surrey and CMMI-EPSRC: Modeling and Monitoring of Urban Underground Climate Change, Grant [EP/T019425/1].

\newpage
\appendix
\section{Multi-building energy system asset specification} \label{app:system-spec}

The specifications of the distributed generation and storage assets for the multi-building energy system simulated in this study are provided in Table \ref{tab:energy-assets}. The battery power capacities are set to be 3 times the mean building electrical load in the test dataset, the battery energy capacities are set to be able to provide 24 hours of that mean electrical load, and the solar PV power capacities are set by assuming 90\% of the roof area of each building is filled by solar panels of power density 0.15 kWp/m$^2$.

\begin{table}[h]
    \centering
    \renewcommand{\arraystretch}{0.75}
    \begin{tabularx}{\linewidth}{c|*{4}{>{\setlength{\baselineskip}{.5\baselineskip}}Y}} \toprule \toprule
        Building & Battery power capacity (kW) & Battery energy capacity (kW) & Battery efficiency (\%) & Solar PV power capacity (kWp) \\ \midrule
        0 & 531 & 4246 & 90 & 461 \\
        3 & 60 & 478 & 90 & 38 \\
        9 & 14 & 108 & 90 & 20 \\
        11 & 342 & 2736 & 90 & 41 \\
        12 & 98 & 780 & 90 & 89 \\
        15 & 203 & 1622 & 90 & 143 \\
        16 & 306 & 2448 & 90 & 120 \\
        25 & 275 & 2198 & 90 & 118 \\
        26 & 1001 & 8006 & 90 & 136 \\
        32 & 157 & 1255 & 90 & 166 \\
        38 & 58 & 461 & 90 & 46 \\
        44 & 1084 & 8674 & 90 & 663 \\
        45 & 1245 & 9962 & 90 & 437 \\
        48 & 622 & 4973 & 90 & 500 \\
        49 & 57 & 458 & 90 & 258 \\
        \bottomrule\bottomrule
    \end{tabularx}
    \caption{Specification of distributed energy assets in simulated multi-building energy system.}
    \label{tab:energy-assets}
\end{table}

\section{MPC planning horizon} \label{app:tau}


The planning horizon used in a MPC scheme determines its ability to effectively arbitrage energy and meet the operational objectives of energy management. Previous studies show that a planning horizon of 24 hours is sufficient when controlling HVAC systems \cite{oldewurtel2012UseModelPredictive}, and recommend 48 hours as a suitable value for systems with distributed solar and storage \cite{thieblemont2017PredictiveControlStrategies}.

\begin{figure}[h]
    \centering
    \includegraphics[width=0.6\linewidth]{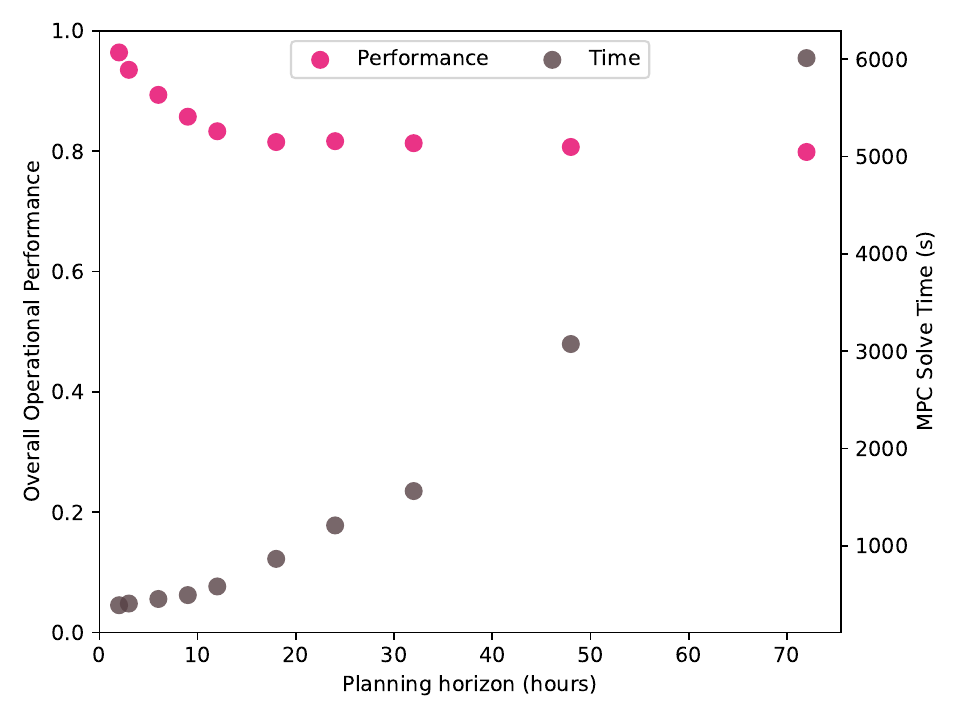}
    \caption{Variation of perfect forecast MPC operational performance and computation time with planning horizon, $T$.}
    \label{fig:planning-horizon}
\end{figure}

To demonstrate the suitability of a planning horizon of $T=48$hrs for the experiments on the test system in this study, the operational performance of the MPC scheme using perfect forecasts is evaluated over varying planning horizon lengths, along with the computational time required to solve the Linear Programs for the simulations, shown in Fig. \ref{fig:planning-horizon}. It can be seen that $T=48$hrs provides near-optimal control performance at a reasonable computation time, and so is a suitable choice for the experiments.

\section{Technical specification of prediction models and training procedures} \label{app:models}


The model parameters and training procedure used for machine learning models have a significant impact on their performance, as well as their computational and data requirements. Hence, they must be selected carefully for the target application to maximise model performance with the available computational resources and data.

\subsection{Simple neural models}

For the simple neural models, implementations from the Pytorch Lighting library \cite{falcon2019PyTorchLightning} were used. The models use an input window of 168 time steps (input dimension) and output a forecast of 48 time steps (output dimension). All models were trained using the Adam optimizer \cite{kingma2017AdamMethodStochastic}, with a batch size of 32 and a learning rate determined using the built-in learning rate tuner.

The Linear model consists of a single linear layer, of size 48, with no activation function used.

The Convolution model uses two 1D convolution layers with channels (5, 1) and kernel sizes (5, 6), respectively, with ReLU activation functions. This is followed by an output MLP that brings the dimensionality to 48.

The ResMLP model has an linear layer with 168 nodes, using a residual connection and ReLU activation. A subsequent output linear layer brings the dimensionality to 48.

\subsection{State-of-the-art machine learning models}

For the complex, state-of-the-art models, implementations from the Pytorch Forecasting library \cite{beitner2024Jdb78Pytorchforecasting} were used. The models take in data from the previous 72 time steps along with known covariate data from the coming 48 time steps, and output forecasts of the target variable for the coming 48 time steps. All models were trained using an early stopping trainer with a patience of 5, for a maximum of 50 epochs, with a limit of 800 training batches per epoch. The learning rate was determined using the built-in learning rate tuner. The size parameters, loss functions, and optimizers used are detailed in Table \ref{tab:complex-model-specs}.

\begin{table}[h]
    \centering
    \renewcommand{\arraystretch}{1}
    \begin{tabularx}{\linewidth}{c|*{3}{>{\setlength{\baselineskip}{.5\baselineskip}}Y}} \toprule \toprule
        Model & TFT & NHiTS & DeepAR \\ \midrule
        Hidden size & 48 & 128 & 64 \\
        Drop-out & 0.1 & 0.1 & 0.1 \\
        Loss & Quantile loss & Quantile loss & Normal distribution loss \\
        Optimizer & Adam & AdamW & Adam \\[1ex]
        & \makecell{Attention head size: 4}
        & \makecell{Weight decay: 0.01\\Backcast loss ratio: 0}
        & \makecell{RNN layers: 3} \\
        \bottomrule\bottomrule
    \end{tabularx}
    \caption{Technical specifications of complex, state-of-the-art models.}
    \label{tab:complex-model-specs}
\end{table}

\section{Data similarity metric} \label{app:similarity-metric}

\subsection{Functional data analysis}

Functional Data Analysis (FDA) is used to analyse and compare the energy usage behaviours across the building electrical load dataset of 15 buildings. In this approach, functional principal components (fPCs) are extracted from the data, such that each data sample can be constructed from an equation of the form,

\begin{equation}
    f(t) = \mu(t) + \sum_{i=1}^n \alpha_i \nu_i(t)
\end{equation}

i.e. the data sample, $f(t)$, is constructed from a linear sum of a mean function, $\mu(t)$, and a weighted sum of $n$ fPCs , $\nu_i(t)$, with weightings or `scores' $\alpha_i$. The mean function and fPCs are the same across all data samples, whereas the weightings are unique to each data sample. This has the benefit that to compare data samples, which are functions of time, it suffices to compare the weightings, a low-dimensional representation of the functional data, that can be analysed using standard statistical techniques.

For this analysis, the datasets are pre-processed into daily time histories, such that each data sample is a 24 hour profile of electricity consumption, starting at midnight. For each building, the training dataset comprises 6 years of data and hence 2191 data samples, and the validation and test datasets are both 2 years and hence 731 and 730 data samples respectively (2016 was a leap year).

\begin{figure}[p]
    \centering
    \includegraphics[width=0.6\linewidth]{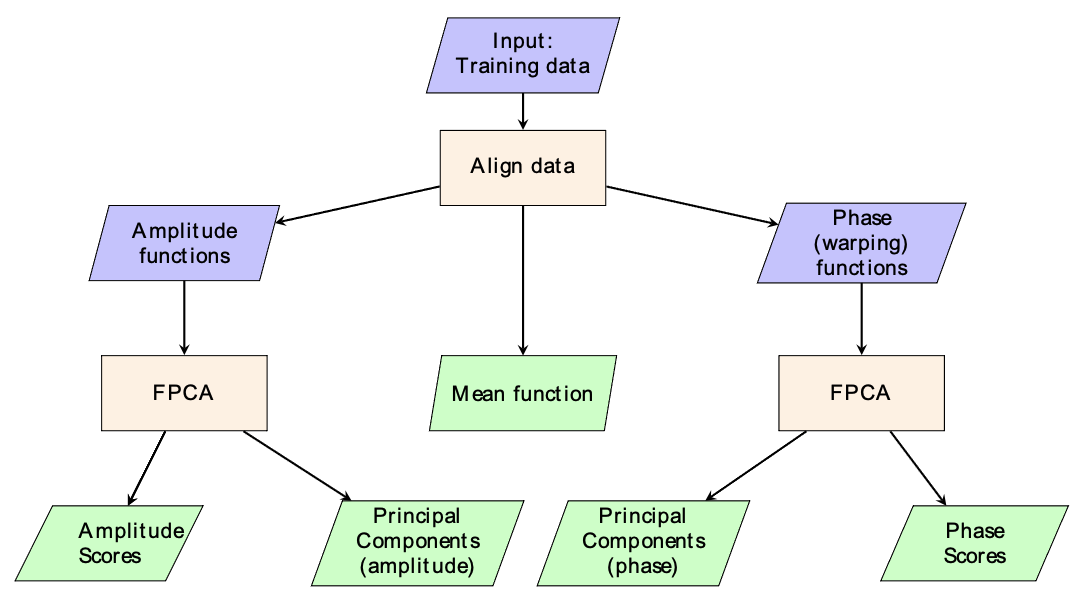}
    \caption{Schematic process for functional principal component analysis (fPCA)}
    \label{fig:fPCA-method}
\end{figure}

The {\color{\hlcolor}functional Principal Component Analysis (fPCA)} approach used involves first aligning the data samples to a common mean, $\mu(t)$. This generates a warping function and an amplitude function for each data sample that describe how the data sample maps to the mean function. The warping function describes the phase relationship, i.e. the variation in time, and the amplitude function describes changes in magnitude. The warping and amplitude functions are then analysed separately and fPCs generated for both. The approach is illustrated schematically in Fig. \ref{fig:fPCA-method}, and full details of the approach are described in (Ward, 2021) \cite{ward2021DatacentricStochasticModel}.

\begin{figure}[p]
    \centering
    \begin{minipage}{.525\textwidth}
        \centering
        \includegraphics[width=0.9\linewidth]{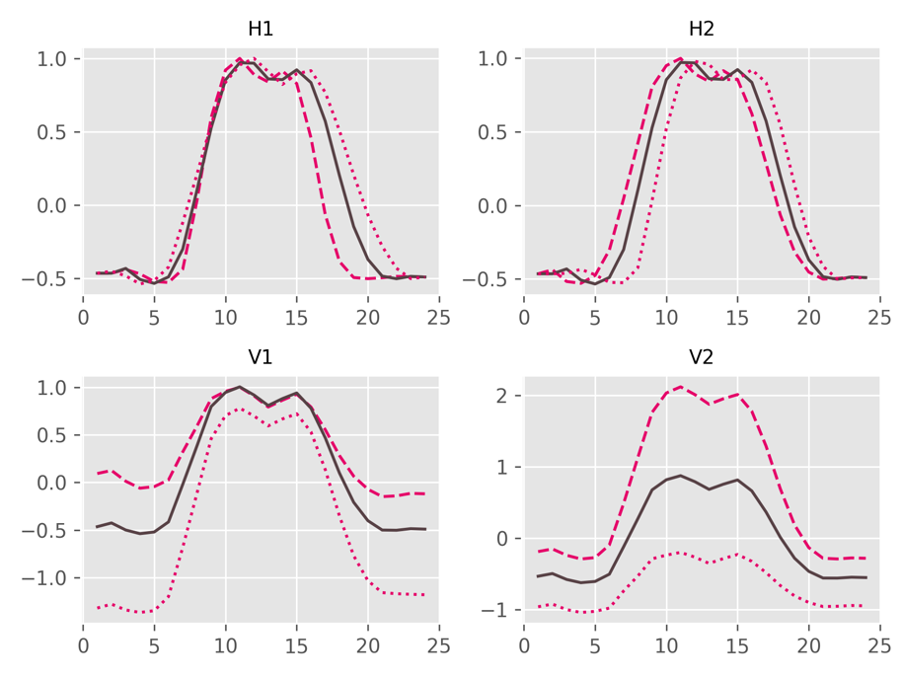}
        \caption{Illustration of first two Phase (H) and Amplitude (V) fPCs. Solid black line shows mean function, $\mu(t)$. Dashed line (-) indicates the impact of a +ve coefficient and dotted line (.), a -ve coefficient.}
        \label{fig:fPC-effects}
    \end{minipage}%
    \hfill
    \begin{minipage}{.425\textwidth}
        \centering
        \includegraphics[width=\linewidth]{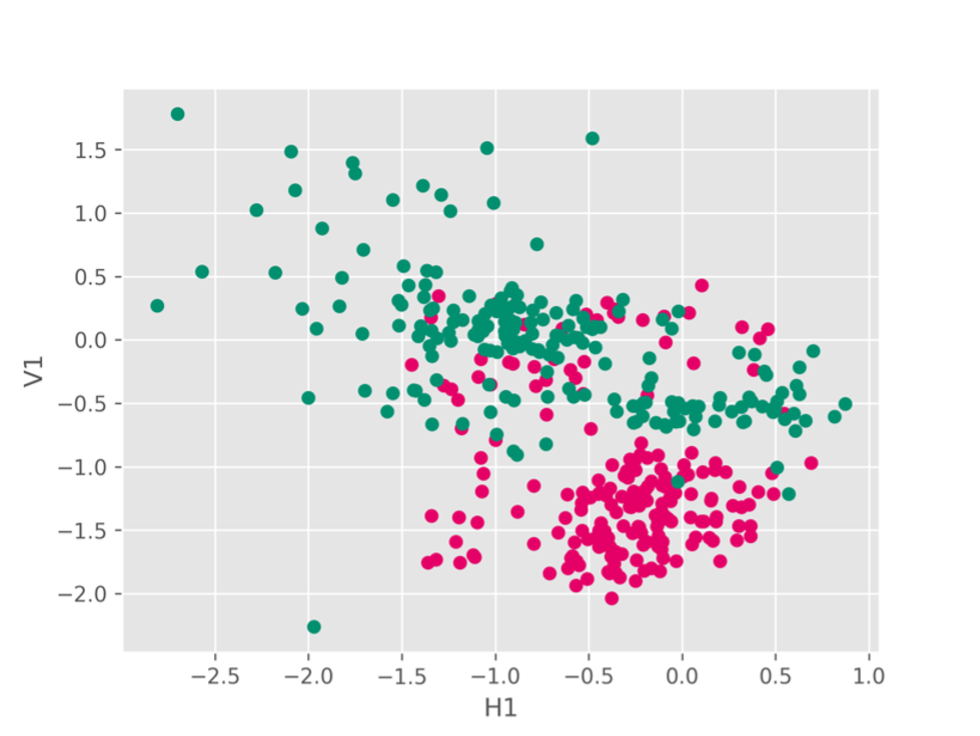}
        \caption{Example fPC coefficient distributions for Buildings 38 (pink) and 49 (green)}
    \label{fig:fPCs-scatter}
    \end{minipage}%
\end{figure}

Fig. \ref{fig:fPC-effects} illustrates the first two phase and amplitude fPCs extracted from the dataset. Fig. \ref{fig:fPCs-scatter} shows distributions of example fPC coefficients for Buildings 38 and 49. The load data for Building 49 exhibits a much lower range than that for Building 38, which results in a more positive distribution of V1 fPC coefficients. In Fig. \ref{fig:fPC-effects} positive coefficients V1 fPC can be seen to reduce the data range.

\begin{figure}[p]
    \centering
    \includegraphics[width=0.4\linewidth]{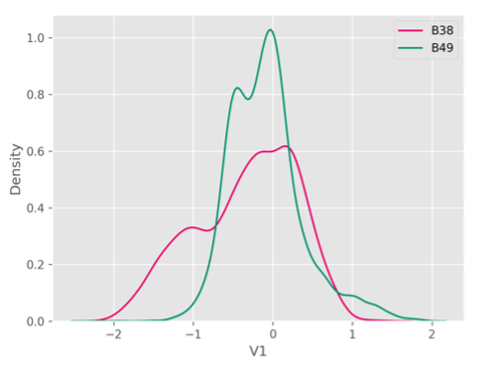}
    \caption{Kernel density plot of V1 scores for Buildings 38 \& 49. Wasserstein distance between two distributions is 0.05.}
    \label{fig:fPC-comparison}
\end{figure}

\subsection{Wasserstein similarity metric} \label{sec:similarity-metrics}

The deconstruction of the data into fPCs and corresponding coefficients enables the use of standard statistical techniques to define measures of similarity between load profiles in the dataset. A load profile similarity metric based on an optimal transport approach is proposed. This metric is based on the approximate Wasserstein distance, or Earth-mover's distance, between the distributions of PC coefficients, which provides a measure of the ease with which one probability distribution can be transformed into the other, and hence is a measure of the similarity of the two distributions. As such, the metric is referred to as the `Wasserstein similarity metric', and is computed using the \href{https://github.com/jeanfeydy/geomloss}{Geomloss} \cite{feydy2024JeanfeydyGeomloss} library. Advantages of this method are that it provides a comparison of the underlying patterns in the load profiles that is not excessively distorted by sequence misalignments as is the case for simple metrics like RMSE, and it allows time series of different durations to be compared.

\begin{figure}
    \centering
    \includegraphics[width=0.6\linewidth]{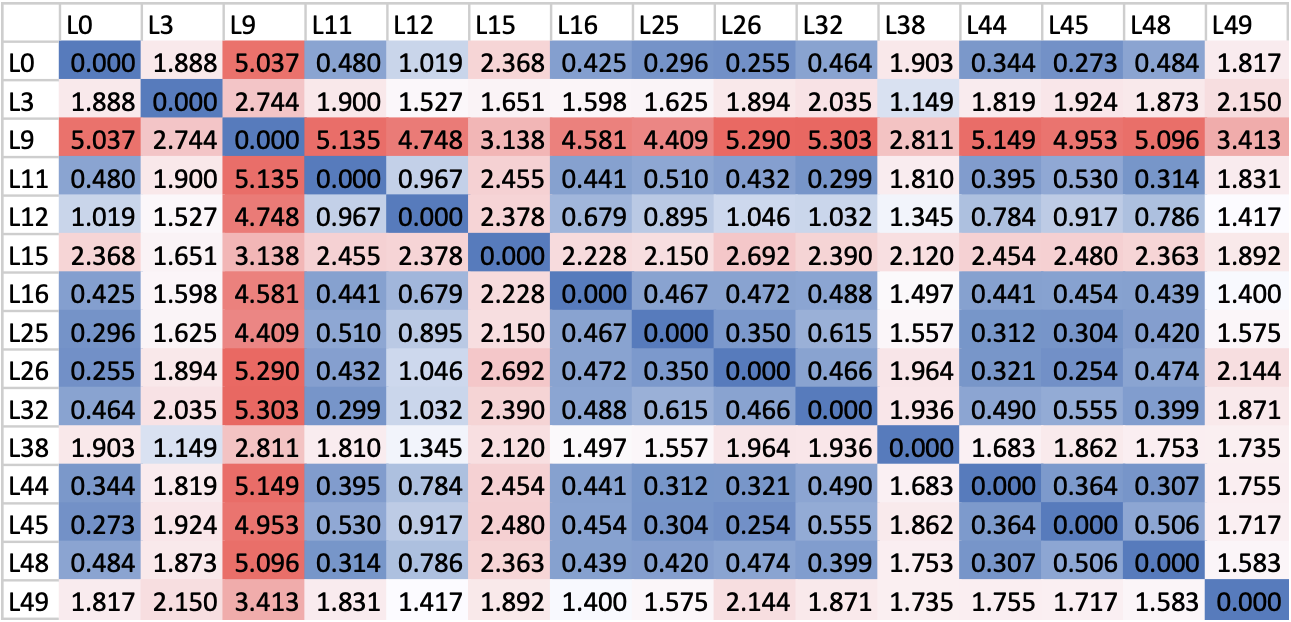}
    \caption{Similarity metric scores between training datasets for pairs of buildings.}
    \label{fig:similarities-buildings}
\end{figure}
\begin{figure}
    \centering
    \includegraphics[width=0.6\linewidth]{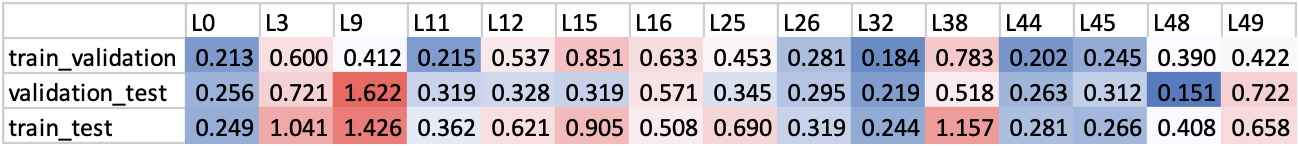}
    \caption{Similarity metric scores between pairs of training, validation, test datasets for each building.}
    \label{fig:similarities-datasets}
\end{figure}

Fig. \ref{fig:fPC-comparison} shows the kernel density distributions of the V1 fPC coefficients for Buildings 38 and 49. The Wasserstein distance between these distributions is 0.05. Similarity metric values are computed for comparison of the training datasets between pairs of buildings, and for the comparison of pairs of training, validation, and test datasets for each building individually. Figures \ref{fig:similarities-buildings} \& \ref{fig:similarities-datasets} provide the results of these similarity metric calculations.

\newpage

\section{Change-point analysis} \label{app:change-points}

Building electrical loads are driven by occupancy behaviour dynamics which span timescales: seasonal (winter heating vs. summer cooling loads), weekly (inhabitance trends due to working patterns, i.e. weekday vs. weekend), and daily (nighttime vs. daytime). The form of these behavioural trends also differs between buildings depending on their usage, e.g. offices and residential buildings have opposing daily trends. Gradual changes can occur in these behavioural trends, for instance due to cultural shifts in working patterns. Alternatively, abrupt changes can be caused by severe disturbances, such as building stock refurbishment, temporary vacancy, or occupant behavioural changes due to change in building use. As such, changes can occur in the seasonal and/or trend patterns of building load time series, and the underlying patterns of the dynamics can differ significantly following disturbances.

This raises the question as to whether all historic measurement data is useful for the training of load prediction models, as the underlying patterns in certain sections of the historic data may differ significantly from the current behaviour of the building load, i.e. be non-representative. Including these sections in the training dataset may poison model training, leading to worse prediction performance.

\begin{figure}[t]
    \centering
    \hspace*{\fill}
    \includegraphics[width=0.425\linewidth]{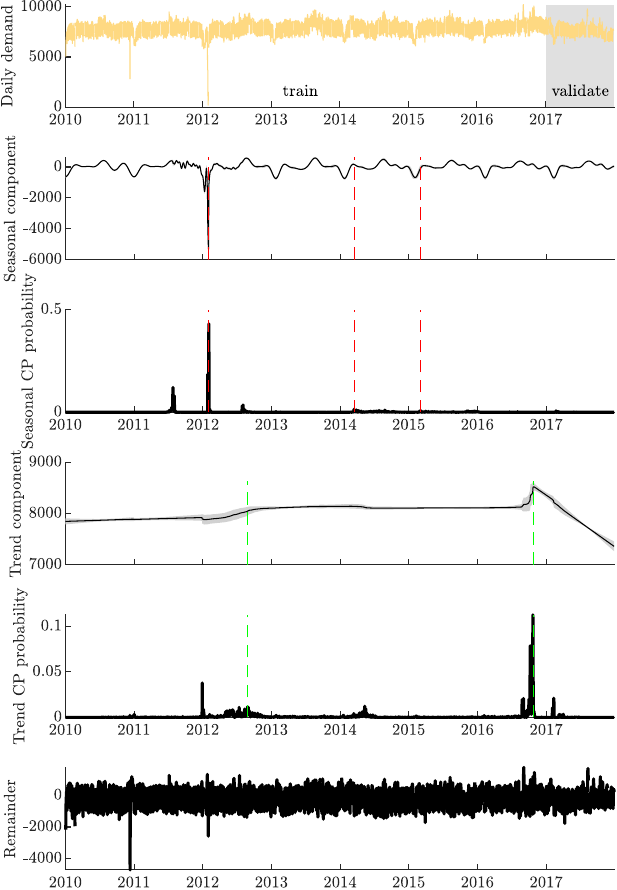}
    \hspace*{\fill}
    \includegraphics[width=0.425\linewidth]{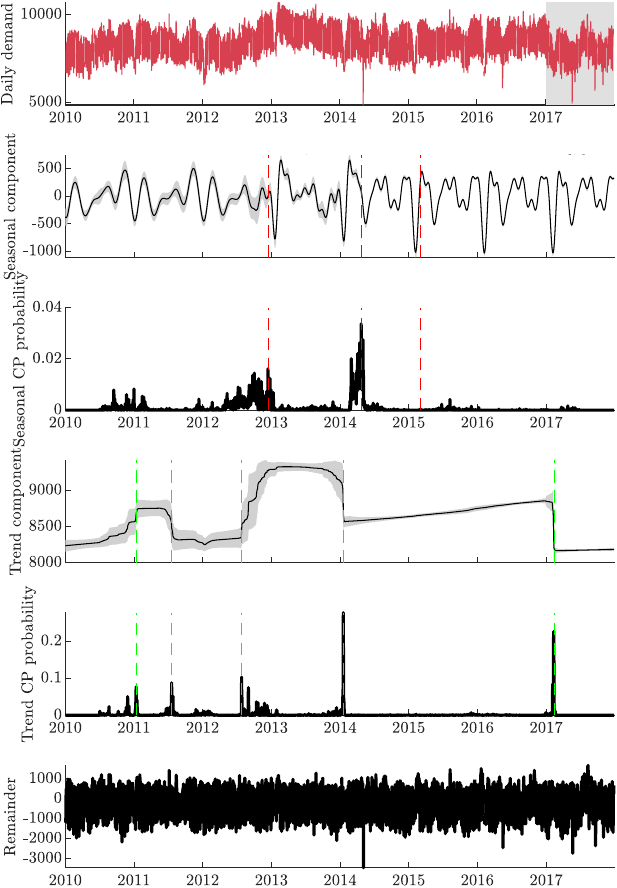}
    \hspace*{\fill}
    \caption{Change-points detected using the BEAST algorithm in load profiles of Buildings 26 \& 44 (top row). Signal is decomposed into a seasonal component (second row), assumed harmonic, inferred change-points indicated by red vertical lines, and a trend component (fourth row), with green lines indicating change-points. Credible intervals of the estimated signals are given by grey envelopes around the individual components.}
    \label{fig:change-points}
\end{figure}

A change-point analysis is conducted on the building electrical load dataset to detect changes in the load patterns in each building. The Bayesian ensemble algorithm ``Bayesian Estimator of Abrupt change, Seasonal change, and Trend'' (BEAST) \cite{zhao2019DetectingChangepointTrend} is used to decompose the load time series and derive non-linear dynamics across multiple timescales, detecting change-points, seasonality, and trends. The value of this ensemble approach is that it quantifies the relative usefulness of individual decomposition models, leveraging all the models via Bayesian model averaging. Additionally, it provides an estimated probability of inferred change-points being `true’ change-points. Results of the change-point analysis for Buildings 26 \& 44 are shown in Fig. \ref{fig:change-points}. The demand profiles are decomposed into seasonal and trend components, and detected change-points for the two components are indicated in red and green respectively. The trend component of Building 44 is found to vary substantially more than that of Building 26, which corresponds with the visual variability of the demand profiles.

\newpage
\bibliographystyle{elsarticle-num} 
\bibliography{Annex37_refs}

\end{document}